\documentclass[aps,prd,twocolumn,showpacs,superscriptaddress,groupedaddress,nofootinbib]{revtex4}  

\usepackage{blindtext}
\usepackage{tensor}
\usepackage{amsmath}
\usepackage{amsfonts}
\usepackage{bm}
\usepackage{color}
\usepackage{graphicx}
\definecolor{darkgreen}{cmyk}{0.85,0.2,1.00,0.2}

\usepackage{amsbsy}
\usepackage{symb}
\usepackage{amsmath,amssymb,subfigure}
\usepackage{graphicx}
\usepackage{epsfig}
\usepackage{color}
\usepackage{ulem}
\usepackage{epstopdf}
\newcommand{\be}{\begin{eqnarray}}

\newcommand{\ee}{\end{eqnarray}}

\usepackage{graphicx}
\usepackage{symb}
\usepackage{epsf}
\usepackage{amsmath}
\usepackage{amssymb}
\usepackage{graphicx}
\usepackage{gensymb}
\def\ok{\Omega_k}

\def\({\left(}
\def\){\right)}

\def\integ0a{\int_0^a}

\def\be{\begin{equation}}
\def\ee{\end{equation}}
\def\bea{\begin{eqnarray}}
\def\eea{\end{eqnarray}}

\def\da_0/do_k{\(-\frac{1}{2} \frac{a_0}{\ok}\)}
\def\de/dE/do_m{\frac{1}{2} \frac{1}{E} \left [ \left ( \frac{a}{a_0} \right ) ^{-3} - f\right ]}

\def\df/da_0{\left( \( \frac{3(1+w_0+w_a)}{a}  \(\frac{a_0}{a}\)^{3(1+w_0+w_a)-1} \times \exp \left \{3w_a\left(\frac{a_0}{a}-1 \right) \right \}\)  +  \(\frac{a_0}{a}\)^{3(1+w_0+w_a)} \times 3\frac{w_a}{a} \exp \left \{3w_a\left (\frac{a_0}{a}-1 \right ) \right \}\right)}

\def\df/do_k{\(\df/da_0\) \(\da_0/\do_k\)}

\def\dfdw0#1{f(#1) 3 \ln(1+#1)}

\def\ok{\Omega_k}

\def\({\left(}
\def\){\right)}

\def\integ0a{\int_0^a}

\def\be{\begin{equation}}
\def\ee{\end{equation}}
\def\bea{\begin{eqnarray}}
\def\eea{\end{eqnarray}}

\def\h0{H_0}

\def\ok{\Omega_k}

\def\w0{w_0}

\def\({\left(}
\def\){\right)}

\def\da{d_A(z)}

\def\ok{\Omega_k}

\def\({\left(}
\def\){\right)}

\def\integ0a{\int_0^a}

\def\bi{\begin{itemize}}

\def\ei{\end{itemize}}
\def\be{\begin{equation}}
\def\ee{\end{equation}}
\def\bea{\begin{eqnarray}}
\def\eea{\end{eqnarray}}

\definecolor{mauve}{rgb}{0.5451,0.2275,  0.3843} 

\begin{document}
\title{Multi-wavelength constraints on the inflationary consistency relation}
\author{P. Daniel Meerburg}
\affiliation{Department of Astrophysical Sciences, Princeton University, Princeton, NJ 08540 USA}
\affiliation{Canadian Institute for Theoretical Astrophysics, University of Toronto, Toronto, Ontario M5S 3H8, Canada}
\author{Ren\'{e}e Hlo\v{z}ek}
\affiliation{Department of Astrophysical Sciences, Princeton University, Princeton, NJ 08540 USA}
\author{Boryana Hadzhiyska}
\affiliation{Department of Astrophysical Sciences, Princeton University, Princeton, NJ 08540 USA}
\author{Joel Meyers}
\affiliation{Canadian Institute for Theoretical Astrophysics, University of Toronto, Toronto, Ontario M5S 3H8, Canada}

\begin{abstract}
We present the first attempt to use a combination of CMB, LIGO, and PPTA data to constrain both the tilt and the running of primordial tensor power spectrum through constraints on the gravitational wave energy density generated in the early universe. Combining measurements at different cosmological scales highlights how complementary data can be used to test the predictions of early universe models including the inflationary consistency relation. Current data prefers a slightly positive tilt ($n_t = 0.06^{+0.63}_{-0.89}$) and a negative running ($n_{t, {\rm run}} < -0.22$) for the tensor power spectrum spectrum. Interestingly, the addition of direct gravitational wave detector data alone puts strong bounds on the tensor-to-scalar ratio $r < 0.2$ since the large positive tensor tilt preferred by the \textit{Planck} temperature power spectrum is no longer allowed. Adding the recently released BICEP2/KECK and \textit{Planck} 353 GHz polarization cross correlation data gives an even stronger bound $r<0.1$. We comment on possible effects of a large positive tilt on the background expansion and show that depending on the assumptions regarding the UV cutoff ($k_{\rm UV}/k_* = 10^{24}$) of the primordial spectrum of gravitational waves, the strongest bounds on $n_t  = 0.07^{+0.52}_{-0.80}$ are derived from this effect.  \end{abstract}

\maketitle

\section{Introduction}

The detection of B-mode polarization on large angular scales by the Background Imaging of Cosmic Extragalactic Polarization II \citep[BICEP2]{BICEP2} collaboration, and its possible primordial origin as the result of relic gravitational waves has invigorated the cosmological community. There has been much debate about the interpretation of the BICEP2 data, and this is an important issue that will be resolved with independent measurements and multi-frequency observations. The recent BICEP2/KECK and \textit{Planck} 353GHz polarized map joint analysis has shown that at least a part of the BICEP2 and KECK signal is due to polarized dust \cite{BICEPKECK2015}. However, more importantly the observations have yielded renewed interest in constraining models of the early universe through their predictions of the spectrum of primordial gravitational waves and the testability of the inflationary scenario. 

Fortunately, at least for a simple class inflationary models in which a single scalar field drove a period of inflation more or less directly preceding the radiation dominated era of hot big bang expansion of our universe, positive predictions can be used to test the inflationary paradigm. The primordial power spectrum of tensor modes is typically parameterized as a power law \citep{Liddle1993}:
\begin{equation}
P_t(k)=A_t(k_*)\left(\frac{k}{k_*}\right)^{n_t} \, , \label{eq:tensor_pk}
\end{equation}
where $A_t$ is the tensor amplitude, $k$ is the wavenumber, $k_*$ is some reference wavenumber, and $n_t$ is the tensor spectral index, or tilt. While there are, in general, deviations from this simple expression, it serves to parameterize a host of scenarios.  For the simplest single field models of inflation there is a relation between the tensor spectral tilt and the tensor-to-scalar ratio $r$, sometimes referred to as the inflationary consistency condition \citep{liddle_lyth}. 

At leading order in slow-roll, this relation is given by \cite{1992PhLB..291..391L,1993PhRvL..71..219C}
\begin{equation}
	r\equiv \frac{A_t}{A_s}=-8n_t \, ,
\end{equation}
where $A_s$ is the amplitude of the scalar power spectrum.  In more complicated models with, for example, a non-trivial sound speed or multiple dynamical fields, this relation is modified to be $r\leq-8n_t$ (see e.g.  \cite{2014arXiv1409.2498P,2014arXiv1407.2621B,palma/soto:2014} for recent discussions of some of these issues). In all models of inflation where the tensor perturbations originate as vacuum fluctuations, the tensor amplitude $A_t$ is set only by the expansion rate which is nearly constant during inflation and always decreases when the null energy condition holds: inflation always predicts a small negative tensor tilt.  The specific value of the tilt $n_t$ may depend on the details of the model, but a negative tensor tilt is a generic prediction of inflation when the null energy condition holds. There are however, other early universe models which make different predictions for the tensor spectrum \cite{2008PhR...465..223L,2004PhRvD..69l7302B,2007PhRvL..98w1302B}. Hence we take a phenomenological point of view and do not restrict our analysis to negative values of $n_t$. If the data provide evidence for a positive tensor tilt, we should further explore alternatives to inflation.  A detection of a positive tensor tilt would suggest at the very least that the primordial gravitational waves were not produced as vacuum fluctuations during inflation\footnote{That being said, counter examples do exist even in the context of inflation \cite{2014JCAP...08..016S,2014arXiv1412.0665M,2012PhRvD..85b3525B}, but fortunately these models tend to make additional predictions for other observables such as non-Gaussianity which would allow additional checks of these scenarios.}.

The aim of this paper is to perform the first {\it joint analysis} of cosmic microwave background (CMB), Laser Interferometer Gravitational wave Observatory \citep[LIGO]{abbott/etal:2007}, and Parkes Pulsar Timing Array \citep[PPTA]{manchester/etal:2013} data to put a constraint on both the tilt $n_t$ and {\it and} the running $n_{t,\rm run}$ of the primordial tensor power spectrum. Previous studies have shown that such combinations of data would improve the available lever arm of the data \citep{smith/peiris/cooray:2006}. BICEP2 data alone yields only a weak constraint on the tensor tilt $n_t$ as the polarization signal of gravitational waves drops rapidly as a function of multipole; hence the CMB alone provides a limited lever arm to constrain the tensor power spectrum. We will show that the constraint on $n_t$ is dominated by the \textit{Planck} temperature power spectrum data, which prefers lower power on large angular scales\footnote{To put any reasonable constraints one has to take a pivot scale that is within the observable window. In this paper we chose the slightly unconventional $k_* = 0.01$ Mpc$^{-1}$. See e.g. Ref.~\cite{2014arXiv1409.6530C} for a discussion of pivot scale in the context of constraining gravitational waves with B-mode polarization.}, resulting in a preference for positive tensor tilt. The additional data extend our lever arm significantly by increasing the scales over which we can measure the gravitational wave energy density by 18-20 orders of magnitude.  Although the constraints at the small scales probed by LIGO and PPTA are only upper limits on the gravitational wave energy density, earlier estimates have shown \citep{Caligiuri2014,Boyle2007} that these experiments are particularly sensitive to the tilt of the primordial tensor mode power spectrum, assuming a detection of a tensor-to-scalar ratio $r \gtrsim 0.01$. The results presented in this paper are relevant for constraining the details of the gravitational wave power spectrum, independent of the current interpretation of the BICEP2 data. We will perform each analysis with and without BICEP2 data, and show that one can get meaningful results even without constraints from BICEP2 on $r$. In addition, we will include the recent released cross analysis data of BICEP2/KECK and the 353 GHz \textit{Planck} polarization data \cite{BICEPKECK2015}. 
%

The paper is organized as follows. In Sec.~\ref{sec:review} we briefly review the relevant physics. In Sec.~\ref{sec:results} we present our results on the bounds on $n_t$ and $n_{t,\rm run}$ from a variety of data combinations, including in Sec.~\ref{sec:discussion} a discussion of the observational constraints derived from the impact of the primordial gravitational wave energy density on the background expansion of the universe not included in the initial analysis. 
We summarize some future prospects and conclude in Sec.~\ref{sec:conclusion}. A summary of the derived parameter constraints using various combination of data sets can be found in Tab.~\ref{table:n_t values}. All bounds quoted in the paper are $95\%$ bounds.  A discussion of the definition of gravitational wave energy density is included in Appendix \ref{appendixA}.

\section{Gravitational wave spectrum and energy density}\label{sec:review}

While measurements of the B-mode polarization of the CMB constrain the tensor mode power spectrum directly, observations from gravitational wave observatories and pulsar timing arrays instead place limits on the energy density of gravitational waves at specific wavenumbers \citep{mtw1973,krauss/white:1992}
\bea
\Omega_\mathrm{GW}(k)&=&\frac{P_t(k)}{12H_0^2}\left[\dot{\mathcal{T}}(\eta_0,k)\right]^2 \, , \label{eq:omgw}
\eea
where $\eta_0$ is the conformal time in the current epoch, $\mathcal{T}(\eta,k)$ is the tensor transfer function, and an overdot refers to a derivative with respect to cosmic time $t$. The transfer function projects the primordial signal  from early to late times and is obtained by solving the Klein-Gordon equation for the tensor fluctuations in an expanding universe\footnote{In general a non-standard thermal history will affect the form of the transfer function.  See for example \cite{Boyle2007,2014arXiv1407.4785K}.}. See Appendix~\ref{appendixA} for a derivation of this formula.

In the case of single field slow-roll inflation, the power spectrum of the tensor modes, $P_t(k),$ is well described by Eq.~\eqref{eq:tensor_pk} with $n_t \simeq 0$. 
The leading deviation from the simple power law Eq.~\eqref{eq:tensor_pk} expected in some models is parameterized by allowing the tilt to vary with scale:
\begin{equation}
P_t(k)=A_t(k_*)\left(\frac{k}{k_*}\right)^{n_t(k_*)+\frac{1}{2}n_{t,\mathrm{run}}\ln(k/k_*)},
\end{equation}
where $n_{t,\mathrm{run}} = dn_t/d\ln k$ is the running of the tensor spectral index and again $A_t$ is the amplitude of the tensor power spectrum.
Single field slow-roll inflation makes a model-dependent prediction for the tensor-to-scalar ratio, i.e. $r\equiv A_t/A_s=16\epsilon$ \citep{Liddle1993}. The (single field slow-roll) inflationary consistency condition also tells us that $n_t = -r/8$, and similar relations can be derived for the running and higher order terms which are generally slow-roll suppressed \citep{Liddle1994}.  As a result, the scale dependence of tensor fluctuations is described by a limited number of parameters, and should be uniquely determined by a limited number of measurements. Despite being slow-roll suppressed, higher order terms can become quite significant in determining the spectrum far from the pivot scale $k_*$ \citep{boyle/etal:2014}.  We are not yet in the data-driven regime, however, and with only upper limits from PPTA and LIGO on the gravitational wave energy density, the data is not yet capable of falsifying the consistency conditions with high significance. Hence, we include the running but leave a treatment of higher order corrections to future work.

In this paper we assume that the power law form of the tensor power spectrum holds all the way up to some fixed ultraviolet cutoff, which we will discuss in more detail below.  Physically, this means that we are assuming an instantaneous transition from the phase of the early universe responsible for producing the primordial spectrum to the phase of radiation domination.  This instant reheating assumption is likely to be violated in a realistic model of the early universe where one would expect modifications to the power spectrum at wavenumbers near the cutoff (see  \cite{2014arXiv1407.4785K} for an excellent review). Other effects, such as the late production of entropy can also modify the spectrum and weaken the constraints derived below. Hence our constraints apply to a phenomenological model with the simplest assumptions: a simple power law primordial spectrum (including the possibility of running), instantaneous reheating, and a standard thermal history.  Any specific model which violates one or more of these features is likely to be subject to different observational constraints.

We use a pivot scale of $k_*=0.01$ Mpc$^{-1}$, particularly because the BICEP2 experiment is insensitive to smaller scales \cite{2014arXiv1409.6530C}. We use an analytic solution for the gravitational wave transfer function valid at late times, which has an accuracy of about 1\% in a flat universe and is relevant for very small scales $k\gg k_{eq}$ and late times $\eta \gg \eta_{eq}$ \citep{WMAP}. 
\begin{figure}[t!] 
   \centering 
   \includegraphics[width=3.5in]{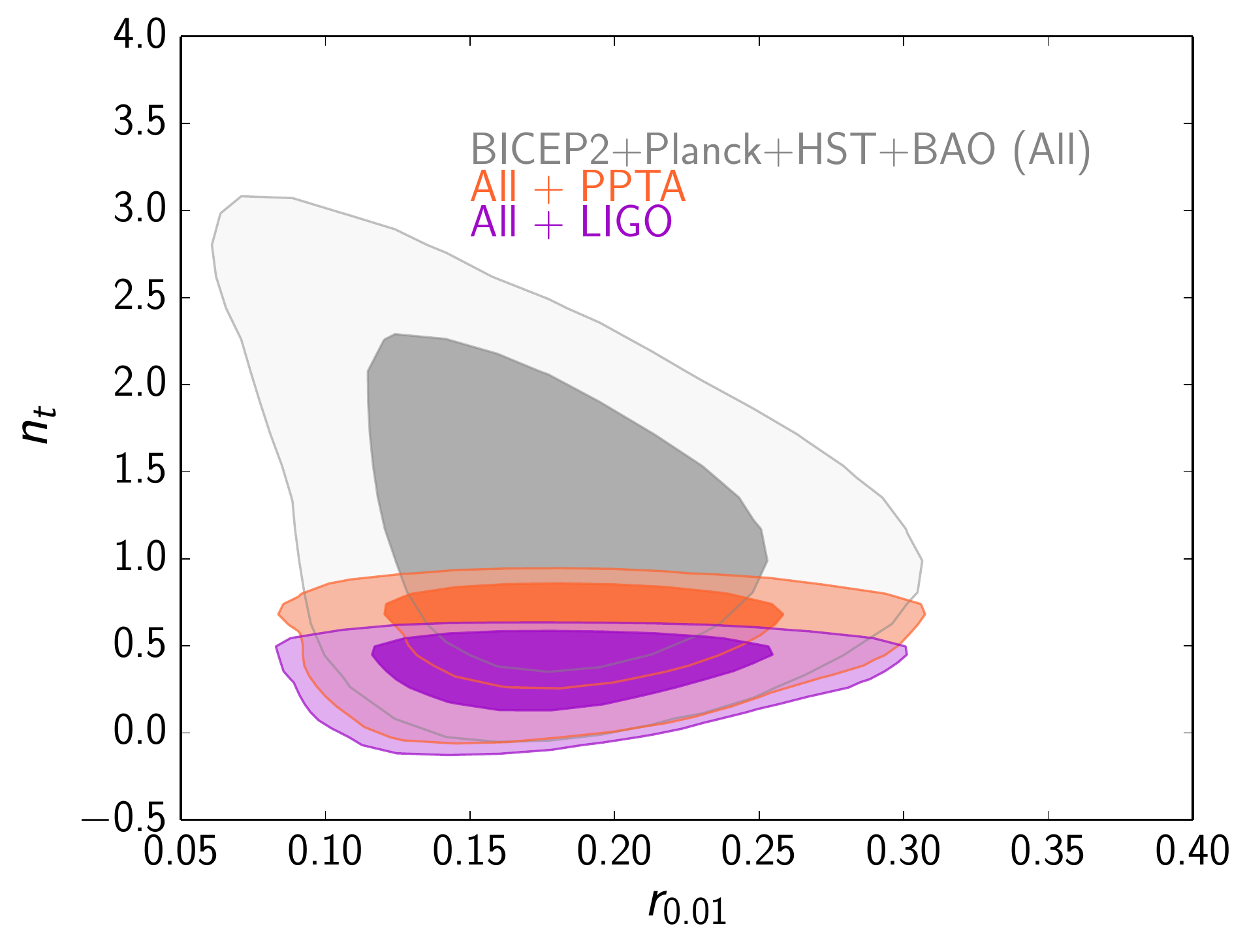} 
   \caption{The marginalized 2D posterior constraints on $n_t$ and $r$. The large grey contour shows the constraints of using BICEP2 in conjunction with CMB data from \textit{Planck} \citep{Planck}, the HST prior on the Hubble constant \citep{HST}, and the constraints on the acoustic scale from BAO \citep{BAO} measurements  (the latter three combined in the name `All'). Adding in data from the PPTA constraints on the gravitational wave energy density improves the constraints three-fold on $n_t$, while the constraint on $r$ is largely unchanged. Using data from LIGO instead tightens this even further, highlighting the constraining power of LIGO on the primordial tensor power spectrum. All contours agree on the value of $r$ announced by the BICEP2 team, confirming that the LIGO and PPTA data are not particularly sensitive to the value of $r$ directly, but they are sensitive to the tilt $n_t$, as shown in Eq.~\eqref{eq:omgw}.}
   \label{fig:nt_r}
\end{figure}
We modified the publicly available Markov Chain Monte Carlo sampler {\tt cosmomc} \citep{cosmomc} and the Boltzmann code {\tt CAMB} \citep{Lewis:1999bs}, by adding a module\footnote{Code will be made publicly available through \url{http://www.astro.princeton.edu/~meerburg/coding/}. } that computes the gravitational wave energy density as a function of wavenumber $k$, which can be converted to frequency using the relationship 
\begin{equation}
f = \frac{k}{2\pi a(\eta_0)} \, .
\end{equation}
The observations of pulsar timing arrays correspond to frequencies of $f \in$($10^{-9}$~Hz, $10^{-7}$~Hz), while gravitational wave observatories probe $f\in$(0.1~mHz, 5000~Hz) \citep{Moore2014}. The constraint from LIGO\footnote{We have approximated the constraints from LIGO by taking a constant upper limit over the frequency range considered. For the purpose of this analysis, this approximation is sufficient.} is $\Omega_\mathrm{GW}<\rm{5.6 \times 10^{-6}}\,(68/H_0)^2$ \citep{LIGO} at $2\sigma$, while the upper limit from PPTA is  $\Omega_\mathrm{GW}<\rm{1.5 \times 10^{-9}}\,(68/H_0)^2$ \citep{PPTA} at $2\sigma$ where $H_0=100\,\,h\, \mathrm{km}\mathrm{s}^{-1}\mathrm{Mpc}^{-1}$.

\begin{figure}[t!] 
   \centering 
   \includegraphics[width=3.35in]{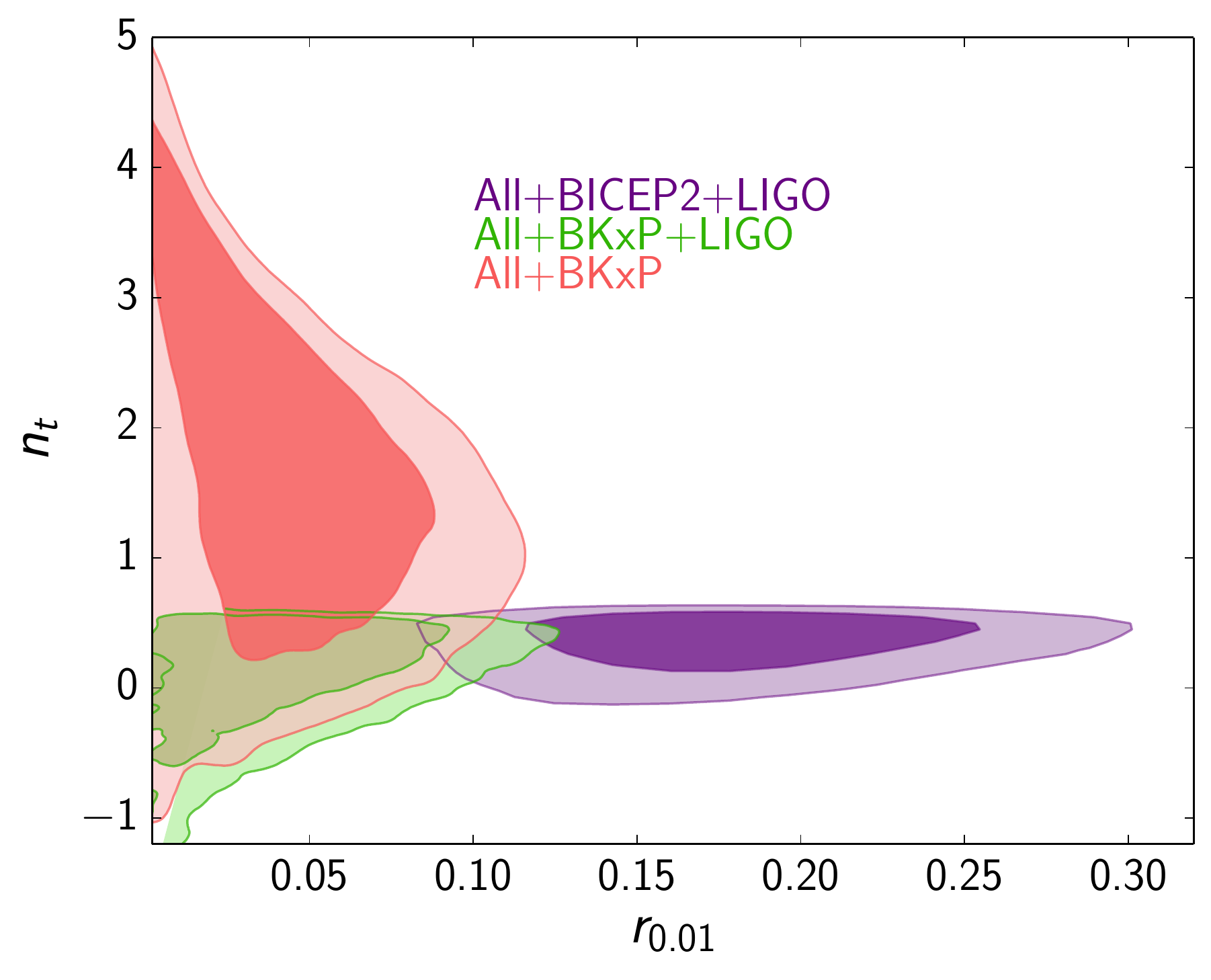} 
   \caption{The marginalized contours in the $r-n_t$ plane comparing BICEP2 to the BKP data. If all the signal in BICEP2 is primordial, we obtain strong bounds on $n_t$, with a $2\sigma$ preference for $n_t>0$. The upper limit is driven by the LIGO constraint. From the BICEP2/KECK cross \textit{Planck} analysis $r$ is lowered and therefore in combination with the constraints from \textit{Planck} TT power spectrum leads to a shift in the peak value of $n_t$. With a small value of $r$, constraints on $n_t$ are weakened. However, including LIGO puts a strong upper bound on $n_t$ as before. }
   \label{fig:BKP_compare}
\end{figure}

\begin{figure}[t!] 
   \centering 
   \includegraphics[width=3.5in]{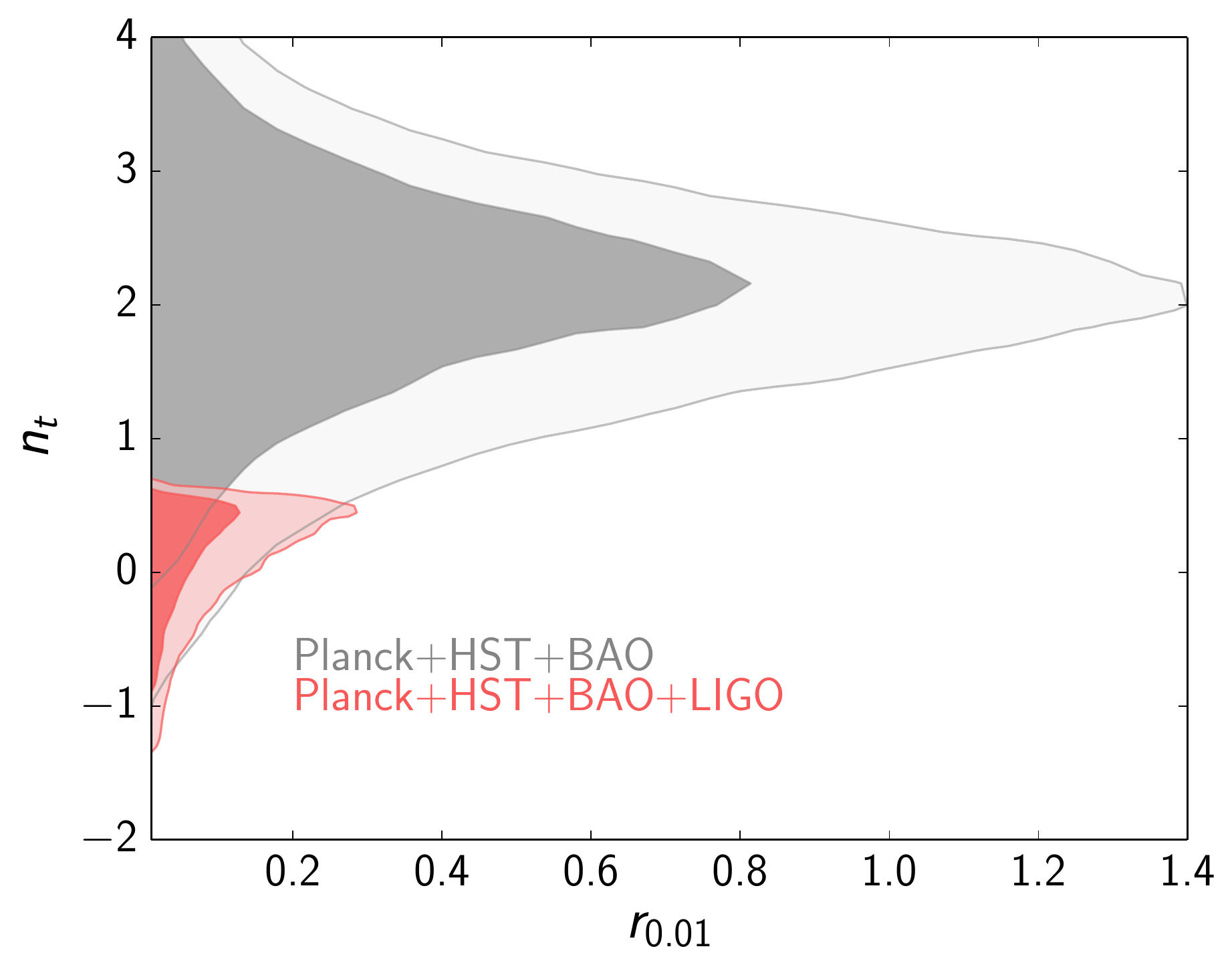} 
   \caption{The marginalized contours in the $r-n_t$ plane without using BICEP2 data. When $r_{0.01} = 0$ there is no constraint on $n_t$ in principle, however \textit{Planck} TT can be best fit with $r_{0.01} = 0.24$, hence there exists some constraint on $n_t$. \textit{Planck} prefers a very blue tensor spectrum, in order to create an artificial running of the temperature data on large scales. From this analysis it is also apparent that \textit{Planck} drives the value of $n_t$, not BICEP2 data. Interestingly, by adding LIGO, such large values of $n_t$ are ruled out, which has the effect that $r$ is significantly constrained $r_{0.01} < 0.2$ at $95\%$.  }
   \label{fig:noBICEP}
\end{figure}

For each value of the gravitational wave energy density, a likelihood is computed based on the upper limits from PPTA and LIGO ($\chi^2=4.0$, corresponding to 2$\sigma$ at 95\% confidence level) from a truncated, logarithmic distribution. We use an analytic solution to obtain the conformal time $\eta$ as a function of scale factor \cite{Sazhin2011}.

\begin{table*}[htbp!]
\begin{center}
\begin{tabular}{|c|c|c|c|c|c|}
\hline
Dataset &\multicolumn{5}{c|}{Parameter} \\ \hline
 & \multicolumn{2}{c|}{$n_t$} &\multicolumn{2}{c|}{$r_{0.01}$}& \multicolumn{1}{c|}{$n_{t,\mathrm{run}}$} \\ \hline
&  Best fit & Mean, 95\% limits& Best fit & 95\% limits & Upper, 95\% limits\\ 
\hline
All & 2.66 &$2.08^{+1.79}_{-1.95}$&0.24 &$<1.0$ & $-$\\
All+LIGO & 0.51 &$0.13^{+0.54}_{-0.75}$ &0.06 & $< 0.2$ &$-$\\
All+PPTA+LIGO (running) & 2.99 &$2.21^{+2.02}_{-1.85}$&1.06 &$ <1.7$ &$< -0.32$\\
All \,$~(N_\mathrm{eff}, k_\mathrm{UV}/k_*=10^{20})$& 0.52&$0.12^{+0.61}_{-0.97}$& 0.01&$ < 0.21$ & $-$\\
All\,$~(N_\mathrm{eff}, k_\mathrm{UV}/k_*=10^{24})$&0.44 &$0.05^{+0.44}_{-0.9}$& 0.01 &$ <0.17$ & $-$\\
\hline
All+BKxP+PPTA &0.65 &$0.32^{+0.71}_{-0.92}$ & 0.04 &$<0.1$ & $-$\\
All+BKxP+LIGO &0.26 &$0.06^{+0.63}_{-0.89}$&0.02 &$<0.1$ & $-$\\
All+BKxP+PPTA+LIGO (running) &3.96 &$1.67^{+2.94}_{-2.77}$&0.02 &$<0.11$ &$ < -0.22$\\
All+BKxP\,$~(N_\mathrm{eff}, k_\mathrm{UV}/k_*=10^{20})$& 0.49&$0.15^{+0.55}_{-0.80}$&0.05 & $<0.09$ &  $-$\\
All+BKxP\,$(N_\mathrm{eff}, k_\mathrm{UV}/k_*=10^{24})$&0.39 &$0.07^{+0.52}_{-0.80}$&0.04 & $<0.09$ & $-$\\
\hline
All+BICEP2 & 1.42& $1.30^{+1.36}_{-1.03}$& 0.18 &$0.18^{+0.1}_{-0.09}$ & $ -$ \\
All+BICEP2+PPTA &0.66 &$0.55^{+0.29}_{-0.49}$&0.22 &$0.19^{+0.09}_{-0.08}$ &$-$\\
All+BICEP2+LIGO&0.56 &$0.35^{+0.20}_{-0.39}$ & 0.18 &$0.18^{+0.07}_{-0.06}$ &$-$\\
All+BICEP2+PPTA+LIGO (running) &0.56 &$1.15^{+1.36}_{-1.21}$&0.19 &$0.19^{+0.1}_{-0.1}$ &$ < -0.05$\\
All+BICEP2\,$~(N_\mathrm{eff}, k_\mathrm{UV}/k_*=10^{20})$& 0.47&$0.4^{+0.09}_{-0.35}$&0.187 & $0.17^{+0.07}_{-0.06}$ & $-$\\
All+BICEP2\,$(N_\mathrm{eff}, k_\mathrm{UV}/k_*=10^{24})$&0.41 &$0.33^{+0.12}_{-0.25}$&0.16 & $0.17^{+0.07}_{-0.06}$ & $-$\\
\hline
\hline

\end{tabular}
 \caption{Summary of parameter constraints using various combinations of data sets. ``All'' refers to \textit{Planck} (2013), WMAP low $\ell$ polarization, HST and BAO data.  In all studied cases a positive $n_t$ is preferred over a negative $n_t$, but the significance is small when you remove any BICEP2/KECK data. Recall that we use a pivot scale $k_* = 0.01$ Mpc$^{-1}$. }
 \label{table:n_t values}
\end{center}

\end{table*}


\section{Results}\label{sec:results}
\subsection{Combining CMB measurements with low-redshift constraints}
In all runs we include temperature data from the \textit{Planck} satellite \citep{Planck} with the polarization prior on the optical depth from WMAP \citep{WMAP}, WMAP low $\ell$ polarization, a prior on the Hubble parameter from the HST key project \citep{HST} and BAO measurements from the Sloan Digital Sky Survey \citep[SDSS]{BAO}. We then include a variety of other probes, including the recent measurements of large-scale polarization from the BICEP2 team \cite{BICEP2}, and constraints on $\Omega_\mathrm{GW}$ from the PPTA or from LIGO data.  In order to be insensitive to lensing effects in polarization data, we restrict the BICEP2 data to the four lowest multipole bins. Last, we include an analysis combining the previous data sets with the BICEP2/KECK and \textit{Planck} cross correlation likelihood \cite{BICEPKECK2015}. Here we fit for the lensing amplitude and the dust contribution. 

Our results are presented in Fig.~\ref{fig:nt_r} and Table~\ref{table:n_t values}. As expected, we get a weak constraint on the 95\% limit value of $n_t$ of $\rm{1.64}<n_t<\rm{2.63}$ when using only BICEP2 data. The constraint on $n_t$ is improved significantly by adding LIGO, lowering the 95\% upper limit by roughly a factor of 5 from 2.63 to 0.59. The mean value we get for the tensor tilt at CL 95\% is $n_t=0.394^{+0.209}_{-0.288}$. The upper limit is a very hard cut, because of the exponential drop off. Our results reflect the knowledge that it is very difficult to measure $n_t$ from CMB data alone (see e.g. Refs.~\cite{Dodelson,boyle/etal:2014}) and small scale GW experiments are necessary to test the inflationary consistency condition. We are also aware that a detection of gravitational waves by LIGO will not be particularly useful for determining the primordial tensor spectrum as the signal on scales to which LIGO is sensitive is dominated by astrophysical sources of GW, such as binary mergers. 
Since current constraints are mainly upper limits on the gravitational wave energy density and any signal which is detected at LIGO is expected to be of non-primordial nature, what we present here are conservative constraints on $n_t$. 


\begin{figure*}[htbp!] 
   \centering 
   \includegraphics[width=5.9in]{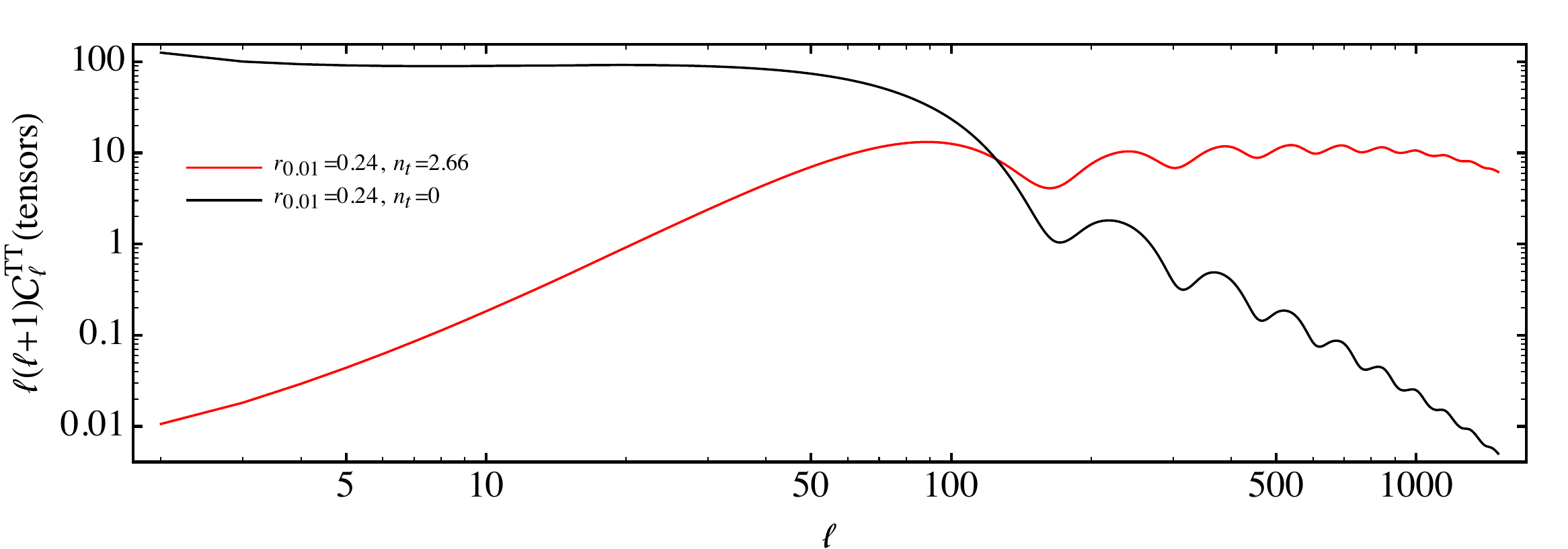} 
   \caption{The contribution from the tensors to the $TT$ power spectrum. A very large positive tilt leads to a rescaling of the spectrum, except on very large scales, where the spectrum is suppressed. This suppression is what \textit{Planck} prefers (though not significantly) over no tensor modes at all.  }
   \label{fig:TensorTTspectra}
\end{figure*}

When we replace the BICEP2 data with the cross data \cite{BICEPKECK2015}, there is a very strong upper bound on the tensor-to-scalar ratio from BICEP2/KECK data, as shown in Fig~\ref{fig:BKP_compare}. We find $n_t = 0.06^{+0.63}_{-0.89}$ and $r_{0.01}<0.1$. 

Even after cross checking \textit{Planck} with BICEP2/KECK data, there is still significant uncertainty in the total fraction due to dust. For completeness, we also considered constraints using only \textit{Planck} temperature data and WMAP polarization, as shown in Fig.~\ref{fig:noBICEP}. In principle, no bound can be put on $n_t$ when $r=0$, but $r=0$ is as likely as any other small value\footnote{We do not encounter serious issues with convergence. An alternative procedure is to constrain $r$ at two different scales and to treat the tilt as a derived parameter. Note that additional scales would be required for constraints on the running and higher order scale dependence.}.  It is interesting to note that \textit{Planck} drives the value of $n_t$, not the BICEP2 data. In fact, allowing $n_t$ to vary freely, leads to the best-fit value $r_{0.01} = 0.24$. We show the contributions from the best-fit spectra to the $TT$ power spectrum at low multipoles in Fig.~\ref{fig:TensorTTspectra}. For very positive tensor tilt, the $TT$ data can be made more consistent with the theory by providing a possibility to lower the power on large scales. This is also the reason why \textit{Planck} alone prefers a large value of $r$ (to increase the overall contribution from the tensors to the $TT$ spectrum) and with a large positive tensor tilt (to mimic suppression of the power on large scales). When LIGO data is added to the likelihood, such preference completely disappears, since $n_t$ can not be too large for most values of $r$. As a result, a very large value of $r$ is no longer allowed and we actually get a constraint on $r$ that is comparable to the constraint from $r$ without adding the tilt and the constraints after adding the BICEP2/KECK and \textit{Planck} cross data. As such, in the scenario that even a larger fraction in the BICEP2 patch is due to dust, simply adding LIGO puts a very strong bound on $r$. That being said, obviously if all measured signal in BICEP2/KECK is due to dust, this puts a very strong bound on $r$. It is interesting to see that even without BICEP2/KECK data one could have put a constraint on $r$ when allowing $n_t$ to vary; adding LIGO tells us that $r < 0.2$, which is consistent with the BICEP2/KECK cross \textit{Planck} data. 
%
%
%
\begin{figure*}[htbp!] 
   \centering 
   \includegraphics[width=5in]{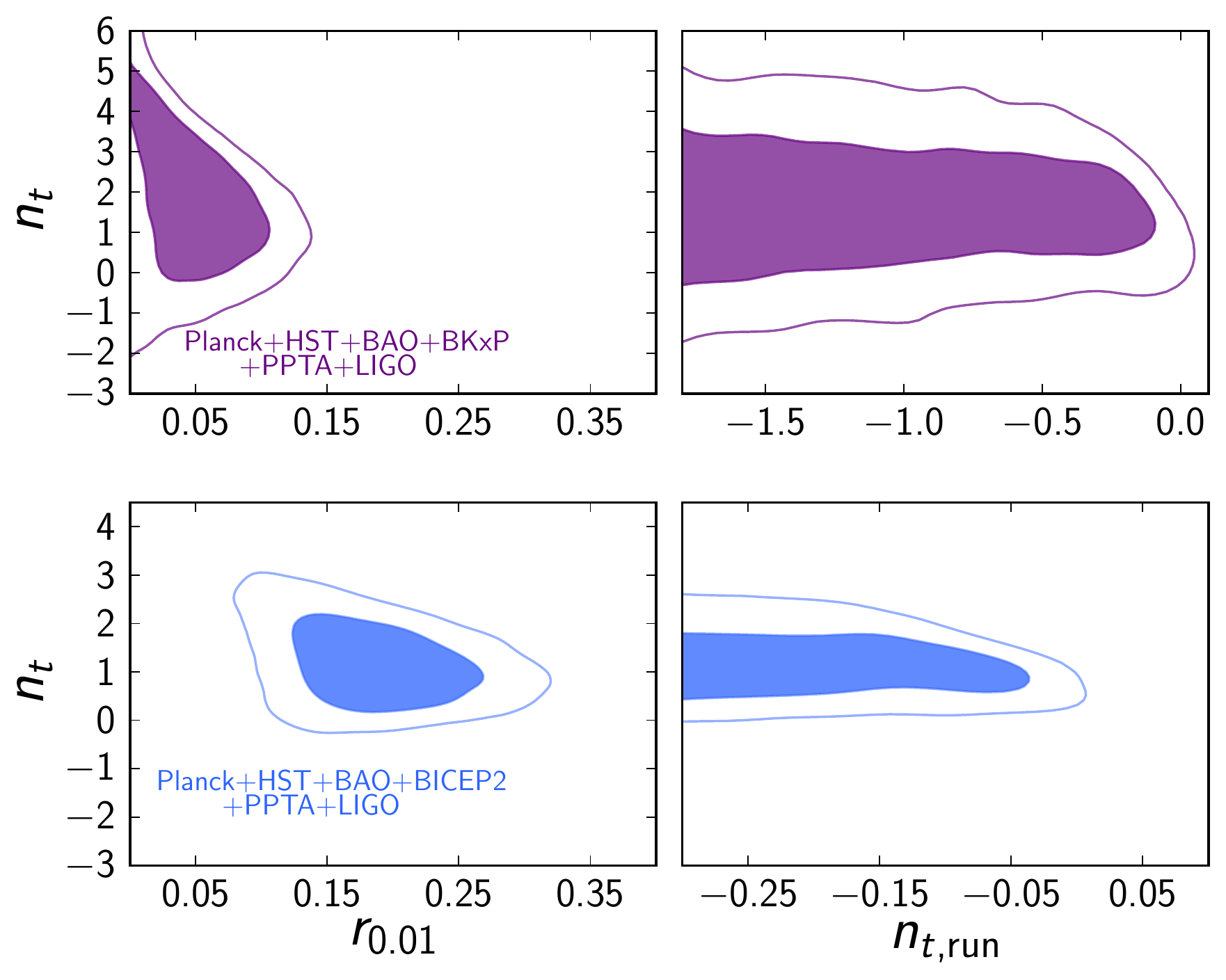} 
   \caption{
Constraints allowing for more freedom in the GW power spectrum. \textit{Left panel -} the marginalized 2D constraints in the $n_t-r$ for the cases including BICEP2 (\textit{bottom}) and BICEP2/KECK cross \textit{Planck} (\textit{top}) data. \textit{Right panel - } constraints on the $n_t - n_{t,\mathrm{run}}$ for the same cases. Allowing for running in the tensor spectral index does not change the constraint on the tilt  significantly, as the errors on the running are large and there is no significant correlation between the two parameters. BICEP2 data improves constraints on $n_t$ as expected. The addition of $n_{t,\mathrm{run}}$ generally weakens the constraint on $r$. Note the different ticks in the figures on the right. }
   \label{fig:DDMLIGO}
\end{figure*}

Next, we consider the running of the tensor tilt. Because the PPTA and LIGO measurements are at different scales we can use both PPTA and LIGO  in our analysis.  The results are presented in the right-hand panel of Figure~\ref{fig:DDMLIGO}. Without small scale constraints on the gravitational wave energy density data are incapable of putting any bound on the $n_{t,{\rm run}}$. As is shown in the right hand panel of Figure~\ref{fig:DDMLIGO}, the running is exclusively confined to negative values and this constraint comes solely from PPTA and LIGO. Naturally, as $n_t$ increases, $n_{t,{\rm run}}$ decreases in order to compensate for the large values of the tilt.  This causes the constraint on $n_t$ to worsen as one moves to lower values of the running. For values above $n_t \sim 3$ (with BICEP2) or $n_t \sim 6 $ (with BICEP2/KECK cross \textit{Planck}), the CMB measurements control the constraint, and the joint data does not allow for larger values of the tilt. Since LIGO and PPTA do not provide lower limits, there is no useful constraints on the lower limit of the running. The constraints on the tensor-to-scalar ratio $r$ are weakened, since running adds another free parameter that allows the fit to improve on large scales in the TT power spectrum. 



\subsection{Gravitational wave contribution to the massless degrees of freedom}\label{sec:discussion}
Our analysis has shown that the inclusion of small scale gravitational wave constraints results in a bound on $n_t \lesssim 0.5$ (at $1\sigma$). Since gravitational waves also contribute to the total radiation energy density of the universe, a consistent analysis must take into account their effect on the background expansion, as was considered in \cite{Boyle2007} and \cite{NeffSmith2006}. The increased radiation energy density alters the peak structure of the CMB as explained in e.g. \cite{2004PhRvD..69h3002B} and \cite{2013PhRvD..87h3008H} and also affects the prediction for the primordial abundance of light elements, see for example \cite{Weinberg:2008zzc}. Rossi et al. \citep{rossi/etal:2014} recently presented constraints from Lyman-$\alpha$ forest measurements in conjunction with a variety of probes.

\begin{figure*}[htbp!] 
   \centering 
   \includegraphics[width=6.8in]{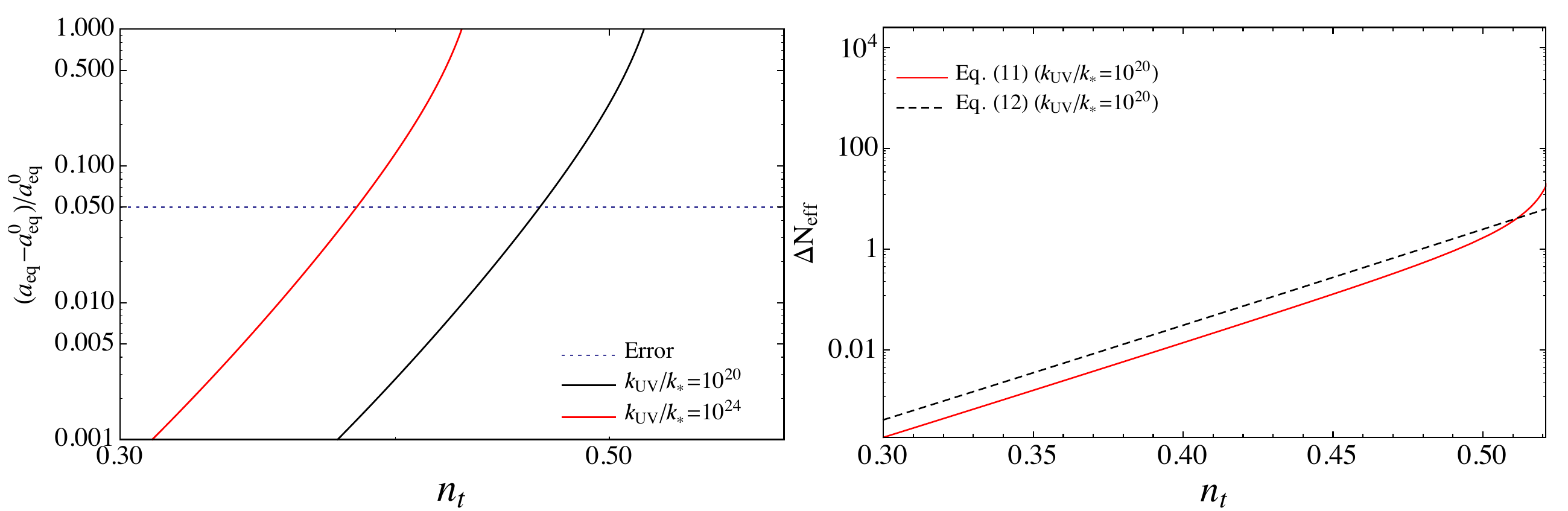} 
   \caption{Left: The relative change in the scale of matter-radiation equality as a function of the tensor tilt $n_t$, assuming $r=0.2$. The simple analytical estimate suggests that $n_t\sim 0.4$ would be modifying the scale of equality at the $5\%$ level, which is close to current bounds from WMAP and \textit{Planck}. Obviously, for smaller $r$, this constraint weakens accordingly. Right: The change in $N_{\rm eff}$ as a function of $n_t$ for $r=0.2$ using the two approximations given in Eqs.~(\ref{eq:Neff}) and (\ref{eq:Neff2}). The difference is very small up until the energy density in gravitational waves becomes of the order of the total energy density in the Universe. Our constraints are derived using Eq.~(\ref{eq:Neff}). A detailed explanation of both curves and how these are derived can be found in the text.  }
   \label{fig:aeq}
\end{figure*}

The energy density of gravitational waves is given by (see Appendix~\ref{appendixA} for details)
\bea
\rho_{\rm GW}  &=& \int_{k_{\rm IR}}^{k_{\rm UV}} d \log k \frac{P_t(k)}{32\pi Ga^2} \left[\mathcal{T}'(k,\eta)\right]^2 \nonumber \\ 
&=& \frac{A_s r }{32 \pi G a^2} \int_{k_{\rm IR}}^{k_{\rm UV}} k dk \left(\frac{k}{k_*}\right)^{n_T} j^2_1(k\eta).
\eea
Here we used that the conformal time derivative of the transfer function during radiation domination is given by $\mathcal{T}'(\eta < \eta_{\rm eq}, k > k_{\rm eq}) = - k j_1(k \eta)$ \cite{Komatsu2006}.  We cannot simply integrate over all wavenumbers, since this integral diverges in the IR for $n_t\leq -4$ and in the UV for $n_t\geq -2$.  Physically, one expects \cite{2005PhRvD..72h3501H} that superhorizon modes should not contribute to the local energy density (note that even this itself is an ambiguous statement in the sense that the total energy density could only be measured after averaging over several wavelengths \cite{mtw1973}) and this behavior is captured by the fact that small wavenumbers make a negligible contribution to the integral.  Furthermore, we do not expect that the physics of the early universe generated primordial gravitational waves at arbitrarily small scales.  The largest conceivable wavenumber at which primordial gravitational waves could have been produced in the early universe is given by the Planck scale at the beginning of the hot big bang expansion of the universe, since modes with larger wavenumbers would have been super-Planckian at some point in the history of the universe.  If we assume that the mechanism which is responsible for solving the flatness and horizon problems is also responsible for generating the primordial spectrum of gravitational waves, then we would expect that the spectrum should extend over a range of scales equivalent to the amount of expansion required to solve those problems.  Below we give results for the case where the spectrum extends only to LIGO scales and also for the case where the UV scale is 60 $e$-folds smaller than the present horizon size.   It is also worth mentioning that although the scaling of the energy for each mode $k$ is equivalent to that of radiation, if there is some comoving IR bound, this scaling is (weakly) broken by the large scale modes which enter the horizon and begin to oscillate. For our purposes, we are only interested in positive $n_t$, for which the UV modes dominate the total energy density. It is natural then to introduce a new variable $q = k/k_{\rm UV}$, $dk = k_{\rm UV} dq$ and $k_{\rm IR}/k_{\rm UV} = \varepsilon$.  In the cases we consider, we can safely take the limit $\varepsilon\rightarrow 0$, but our results will still depend on the choice of $k_\mathrm{UV}$, which we leave unspecified for the moment. 

We can then perform the integral for positive $n_t$
\bea
&k_{\rm UV}^2 \left(\frac{k_{\rm UV }}{k_*}\right)^{n_t}  \int_{0}^{1} q^{n_t +1} dq j^2_1(q k_{\rm UV} \eta)  \nonumber \\
 &\qquad=\left(\frac{k_{\rm UV }}{k_*}\right)^{n_t}  \frac{1}{2 n_t} \frac{1}{\eta^2} +  \mathcal{O}(1/(k_{\rm UV}\eta)) \, .
\eea
Hence for the  gravitational wave energy density we find (up to corrections of order $\varepsilon$)
\begin{equation}\label{rhoGW}
	\rho_{\rm GW} (\eta) \simeq \frac{A_s r}{32 \pi G}  \left(\frac{k_{\rm UV }}{k_*}\right)^{n_t}  \frac{1}{2 n_t} \frac{1}{(a\eta)^2}. 
\end{equation}
It is useful to relate this energy density to quantities which may be more familiar (for a similar analysis, see \cite{Maggiore:1999vm}).  Deep in the radiation-dominated era, after big bang nucleosynthesis (BBN) but well before matter-radiation equality, total energy density of the universe is given by
\begin{align}
	\rho_{\mathrm{tot}}&=\rho_\gamma+\rho_\nu+\rho_{\mathrm{GW}} \nonumber \\
	& \equiv \rho_\gamma\left(1+\frac{7}{8}\left(\frac{4}{11}\right)^{4/3}N_\mathrm{eff}\right) \, ,
\end{align}
where $N_{\mathrm{eff}}=3.046$ in the standard model in the absence of gravitational waves \cite{Mangano:2005cc}.  Using the fact that during this period, $1/(a\eta)^2=H^2=8\pi G \rho_\mathrm{tot}/3$, and also assuming that the gravitational waves are not the dominant source of energy density, we can write the fraction of total energy density in gravitational waves as
\begin{align}
	\frac{\rho_{\mathrm{GW}}}{\rho_\mathrm{tot}}&=\frac{A_s r}{24n_t}\left(\frac{k_\mathrm{UV}}{k_*}\right)^{n_t} \nonumber \\
	&=\frac{\frac{7}{8}\left(\frac{4}{11}\right)^{4/3}\left(N_\mathrm{eff}-3.046\right)\rho_\gamma}{ \left(1+\frac{7}{8}\left(\frac{4}{11}\right)^{4/3}N_\mathrm{eff}\right)\rho_\gamma} \,
\end{align}
which can be solved for $N_{\mathrm{eff}}$
\begin{equation}\label{eq:Neff}
	N_\mathrm{eff}=\frac{\frac{8}{7}\left(\frac{11}{4}\right)^{4/3}\left[\frac{A_s r}{24n_t}\left(\frac{k_\mathrm{UV}}{k_*}\right)^{n_t}\right]+3.046 }{1-\left[\frac{A_s r}{24n_t}\left(\frac{k_\mathrm{UV}}{k_*}\right)^{n_t}\right]} \, .
\end{equation}
This expression breaks down when the gravitational wave energy density becomes comparable to the total energy density of the universe.  A fully consistent analysis for larger values of the gravitational wave energy density would require deriving a new transfer function which does not assume radiation domination, but instead incorporates the backreaction of the gravitational waves on the expansion history.  Luckily, the data is sufficient to constrain the gravitational wave energy density to a level significantly below the regime where this presents a problem.  An alternative approximation for $N_\mathrm{eff}$ which does not have a singularity can be found by neglecting the contribution of $\rho_\mathrm{GW}$ to $H$ in Eq.~(\ref{rhoGW}) which gives the same result as the first order Taylor expansion of Eq.~(\ref{eq:Neff}) about $\rho_{GW}/\rho_{\mathrm{tot}}=0$,
\begin{equation}\label{eq:Neff2}
	N_\mathrm{eff}\approx3.046+\left(3.046+\frac{8}{7}\left(\frac{11}{4}\right)^{4/3}\right)\frac{A_s r}{24n_t}\left(\frac{k_\mathrm{UV}}{k_*}\right)^{n_t} \, .
\end{equation}
Both Eq.~(\ref{eq:Neff}) and Eq.~(\ref{eq:Neff2}) should only be trusted in the regime $\rho_\mathrm{GW}/\rho_\mathrm{tot}\ll 1$.  We use Eq.~(\ref{eq:Neff}) in our analysis since it gives more conservative constraints on $n_t$, though the difference is small with current data (see right panel of Fig.~\ref{fig:aeq} for a comparison of these approximations in the parameter range of interest). 

Since the energy density of gravitational waves redshifts as radiation (so long as we can neglect the contribution of the long wavelength modes) the relative energy density of gravitational waves compared to that of photons, and thus the gravitational wave contribution to the value of $N_\mathrm{eff}$, remains constant for all times of interest.  

In order to estimate the size of the effect of the additional radiation energy density, we will calculate the change in the redshift of matter-radiation equality.  At the instant of matter-radiation equality, we have by definition 
\be
\rho_m(a_\mathrm{eq}) = \rho_r(a_\mathrm{eq}) \, .
\ee
This can be written as an equation for $a_\mathrm{eq}$ as follows
\begin{align}
	\frac{3H_0^2\Omega_m}{8\pi G}&\left(\frac{1}{a_\mathrm{eq}}\right)^3 \nonumber \\
	&= a_\mathcal{B}T_{\gamma,0}^4\left(1+\frac{7}{8}\left(\frac{4}{11}\right)^{4/3}N_\mathrm{eff}\right)\left(\frac{1}{a_\mathrm{eq}}\right)^4 \, ,
\end{align}
where $a_\mathcal{B}$ is the radiation constant, $T_{\gamma,0}$ is the CMB temperature today, $N_\mathrm{eff}$ is given by Eq.~(\ref{eq:Neff}), and we have fixed $a_0=1$.
 
In order to place constraints on $r$ and $n_t$ we have to make an assumption about the ratio $k_{\rm UV}/k_*$. In our analysis of the data we chose a pivot scale $k_* = 0.01$ Mpc$^{-1}$. For LIGO we do not have to make an assumption on the UV cutoff, however we do assume the scaling is power law all the way up to $k_{\rm LIGO}$. The minimum for this is ratio would then be $k_{\rm UV}/k_* \simeq 10^{20}$, while if we assume that the power law spectrum extends over $\sim60$ $e$-folds\footnote{This value is chosen not due to some particular mechanism for generating the gravitational waves in the early universe, but instead to roughly correspond to the maximum amount of hot big bang expansion.} the ratio is given by $k_{\rm UV}/k_* \simeq 10^{24}$.  We estimate the contribution of primordial gravitational waves to the total energy density at the time of equality between matter and the relativistic components of the universe. Shifts in the redshift of matter-radiation equality by roughly a percent will be constrained by the CMB power spectrum, since this the accuracy with which the redshift of equality is measured. In Fig.~\ref{fig:aeq} we plot the relative change of the $a_{\rm eq}$ as a function of $n_t$ for these two numbers.  It can be seen from this estimate that even for the lower value of $k_\mathrm{UV}$, the observational constraint from the modified expansion history will be at a level similar the best constraint from the CMB+LIGO data. 

For consistency we include this effect in our CMB analysis. We add the gravitational wave contribution to the total the massless degrees of freedom in $N_{\rm eff}$ as computed above. The way the sampler usually scans the likelihood is to sample the densities and the optical depth independently (slow parameters) of the primordial parameters (fast parameters); this scheme obviously does not work in case one of the densities depends explicitly on the primordial parameters. For that reason we changed the sampling. Unfortunately this makes the scanning of the likelihood slower, as one adds three additional parameters to the slow part of the code (namely the computation of the radiation transfer functions). In order to speed up the convergence, we therefore fix the foregrounds to their best fit values. 

\begin{figure*}[htbp!] 
   \centering 
   \includegraphics[width=6.5in]{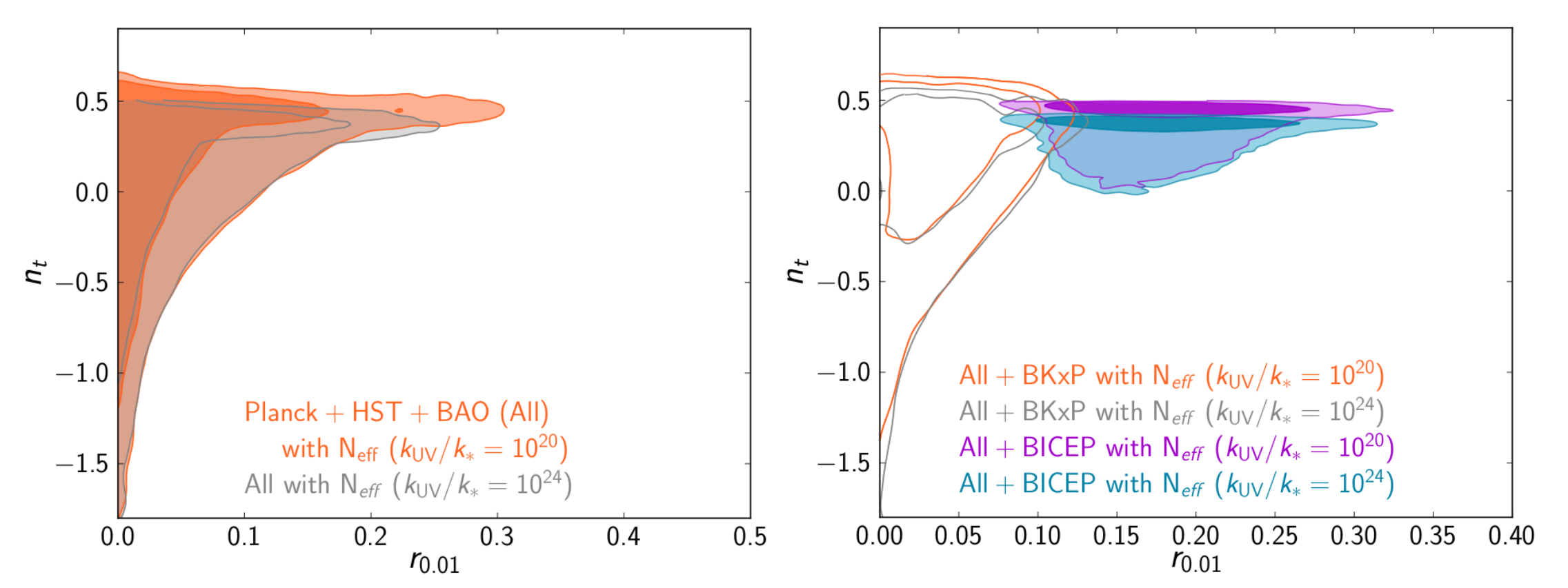} 
   \caption{The constraints on the gravitational wave spectrum when adding in the change in $N_{\rm eff}$ from gravitational waves. Allowing for the $N_{\rm eff}$ contribution cuts off the distribution on $n_t$ to more positive values. Adding in BICEP2 data again moves the constraints to more positive values of $r$, as expected. When considering the BKxP data in addition to \textit{Planck}, the upper limits on $n_t$ are hardly affected as can been see in the figure on the right. We also show the effect of changing the cut-off scale from $k_*/k_{\mathrm{UV}}=10^{20}$ to $k_*/k_{\mathrm{UV}}=10^{24}.$ The overall effect is to limit the upper bound on $n_t$. }
   \label{fig:nt_r_neff}
\end{figure*}

\begin{figure*}[htbp!] 
   \centering 
   \includegraphics[width=6.5in]{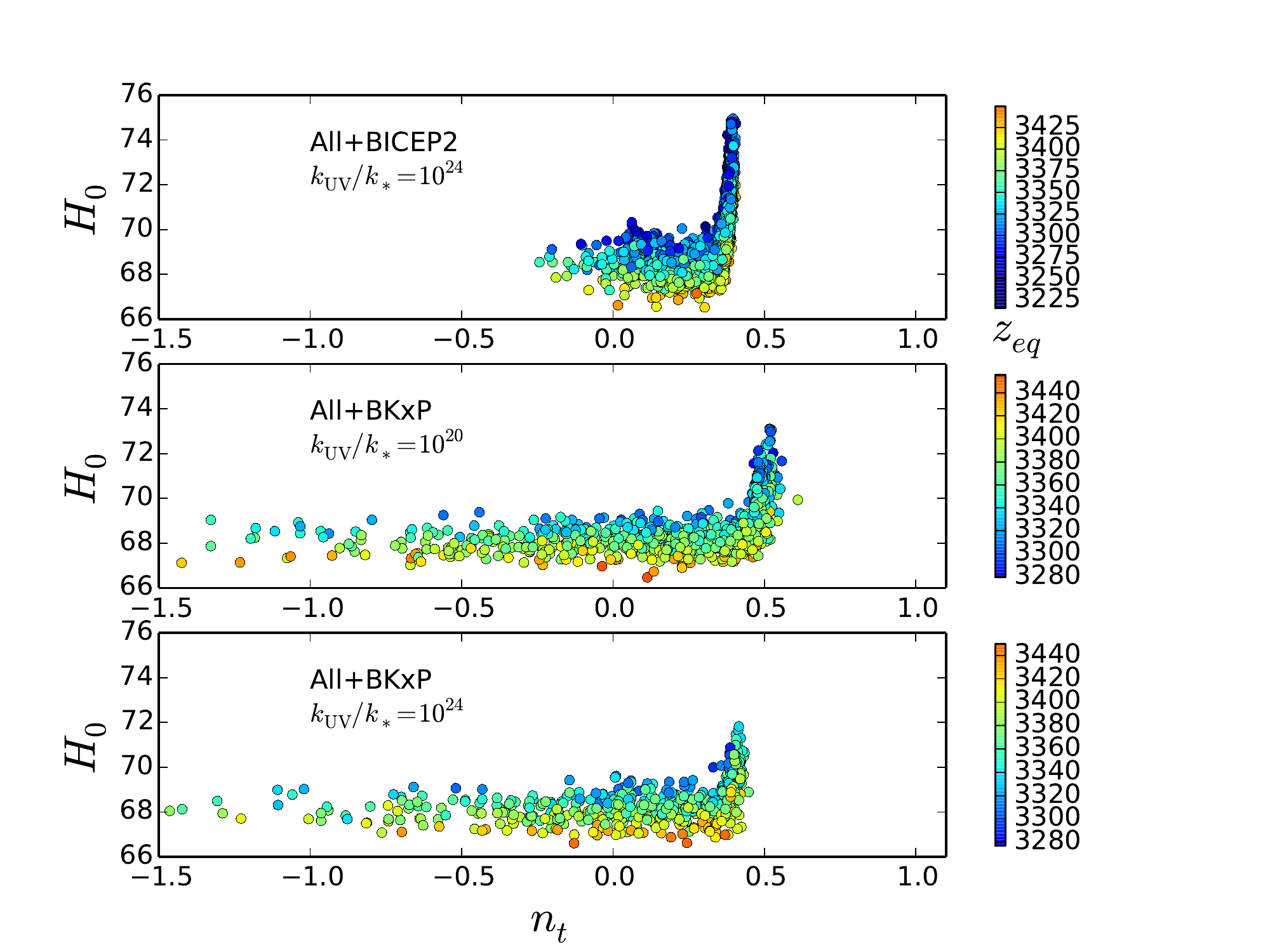} 
   \caption{A scatter plot showing the relation between the tensor tilt $n_t$, the Hubble rate $H_0$ and the redshift of equality $z_{\rm eq},$ including the contribution from gravitational waves to $N_{\rm eff}$. The \textit{top} panel shows the case where the data are combined with BICEP2, while the \textit{middle} and \textit{lower} panels replace the BICEP2 data with the BICEP2/KECK-\textit{Planck} cross spectrum for two different values of the cutoff. In all cases, $z_{\rm eq}$ is fairly well constrained, hence in order to accommodate large $n_t$ (and hence a large $N_{\rm eff}$, $H_0$ is increased by adding more matter to the universe. Because of the steep increase of $N_{\rm eff}$ as a function of $n_t$ there is a very strong turn in degeneracy between $n_t$ and $H_0$. There is a shift in the upper bound of $n_t$ allowed by BICEP2 compared to the cross spectrum, and this limit also shifts with different values of the cut off, as can be seen by comparing the middle to lower panels of the figure.
   \label{fig:scatter_zeq}}
\end{figure*}

The approximations we have used are valid for a positive tensor tilt, and only for small values or negative values of the tilt would we have to worry about IR corrections. However, from the analytical analysis, the background will not be affected in that limit and hence the total energy density from gravitational waves will be negligible. The code does not compute a correction to $N_{\rm eff}$ when $n_t<0$. For very large values of $n_t$, Eq.~\ref{eq:Neff} breaks down predicting negative values of $N_{\rm eff}$, and so in that case, we replace $N_{\rm eff}$ with a very large number. The results are shown in Fig.~\ref{fig:nt_r_neff} for two assumptions about the UV cutoff; one in which the cutoff is set at the LIGO scale and one in which we assume the maximal cutoff $k_{\rm UV}/k_* \simeq 10^{24}$. The constraints are summarized in Tab.~\ref{table:n_t values}. Compared to our analytical estimate, we find the constraints to be slightly weaker. Although $z_{\rm eq}$ is constrained quite strongly, $H_0$ and the matter density tend to vary as $n_t$ increases in order to keep $z_{\rm eq}$ fixed, which therefore gives a weaker constraint than our naive prediction. This is shown in Fig.~\ref{fig:scatter_zeq}. The strongest bound, as expected, is derived when we include BICEP2 data (which drives $r$ to large values) and  $k_{\rm UV}/k_* \simeq 10^{24}$. We find $n_t = 0.33^{+0.12}_{-0.25}$. This stronger bound is driven by a non-zero $r$.  Without the addition of BICEP2 data constraints are weaker and are found to be remarkably close to LIGO bounds. 

The change in the expansion history due to the gravitational wave background energy density also affects BBN. The presence of extra massless degrees of freedom during BBN increases the primordial Helium fraction by increasing the expansion rate and thus decreasing the time during which free neutrons can decay before becoming bound into light nuclei \cite{Weinberg:2008zzc}. In \cite{Brandenberger2007} and recently in \cite{Gerbino2014} it was claimed that BBN puts an upper limit (with unknown confidence) on $n_t \leq 0.15$. However, this constraint is derived under the assumption that $k_{\rm UV}/k_* \simeq 10^{61}$, which would suggest the power law spectrum of gravitational waves extends over more than 120 $e$-folds and that many modes would have certainly been super-Planckian during the hot big bang expansion of the universe. Without a consistent theory of quantum gravity, it is unclear how to treat the evolution of such high energy gravitons. If one repeats the analysis with more sensible numbers, we find $n_t \leq 0.4$ for $r = 0.2$, in line with \cite{Boyle2007} and \cite{2014arXiv1407.4785K}. This bound again is very close to the bound from LIGO and from the CMB, and for completeness it should be taken into account. However, we expect based  on a very similar analysis in \cite{NeffSmith2006} that the modified expansion history leads to effects for which the CMB provides a stronger constraint, given the same $\rm UV$ cutoff. That being said, the BBN constraint depends on the value of both $Y_p$ and $\Omega_b$ and it would be possible to include this fully self consistently inside CAMB. In this analysis we have fixed the value of $Y_p = 0.24$,  and we leave a full treatment including these effects to future work.

When considering the constraints derived from the contribution of gravitational waves to the massless degrees of freedom, there are two important things to remember. Firstly, the total energy density in gravitational waves depends on the entire primordial spectrum, which we have assumed here to be a power law all the way up to some UV cutoff.  A more gradual transition to the reheating phase, or production of additional gravitational waves during reheating could alter the constraints derived here.
In addition, since the constraints on modified expansion history are from `indirect' measures, the effect of gravitational waves would be degenerate with modifications to the neutrino sector or the addition of dark radiation which would have the same phenomenological effect on the measured CMB \cite{NeffSmith2006}. As such, this constraint is less robust than actually putting upper bounds on direct gravitational wave detection. In our analysis we have assumed that beyond photons and three families of standard model neutrinos, all of the radiation energy density is made up of gravitational waves, with no additional massless degrees of freedom making a significant contribution to the energy density of the universe. 

\section{Conclusion}\label{sec:conclusion}

In this paper we performed the first joint analysis of CMB data and late time measures of gravitational waves. This study is interesting for the following reasons. First, a combined analysis allowed us to put a bound on both the tilt {\it and} the running of the primordial tensor power spectrum. Secondly, even without BICEP2 data or the BICEP2/KECK and \textit{Planck} cross data, we can put a bound on the tensor-to-scalar ratio $r$ that is close to the constraint without allowing for the tensor tilt to vary. This is especially important given the recent conclusion that dust could be responsible for all of the detected B-mode power. While additional sources of gravitational waves could explain the detection of $r$ (even without considering dust), this gives interesting bounds on the energy scale of inflation \citep{ozoy/etal:2014}, of the detailed perturbation physics in this early epoch \citep{palma/soto:2014}, and of other early universe physics \citep{2014arXiv1412.0407L}.  Thirdly, it has shown that all possible constraints on the tilt (and the running) are of the same order, and a full analysis should consider all these effects at the same time.  

Current data prefers the tilt to be positive and the running to be negative. The posterior parameters ranges in the combined analysis are reduced by factor 5 for the tilt and infinitely for the running (there is no constraint from the CMB). As expected, the data is not yet capable of putting the inflation consistency condition to the test, but this work suffices as a confirmation that such an analysis is possible and that the combined data is already much more constraining than the CMB alone. A true test of the inflation consistency condition requires improved CMB data and measurement of gravitational waves at frequencies that are not dominated by foregrounds. 

We also showed that for large positive values of the tensor tilt, one expects the background evolution to be modified, altering the peak structure of the CMB. There are some caveats in this analysis, since it requires an explicit assumption about the UV cutoff of the tensor spectrum and one could in principle introduce other ingredients with degenerate cosmological effects, i.e. this measure is not as `clean' as direct constraints obtained with gravitational wave detectors. For that reason, one should be careful interpreting these results. Although analytical results indicated that the constraints would be better, performing the actual analysis showed that parameter degeneracies lead to bounds on the tilt that are slightly worse ($k_{\rm UV}/k_* = 10^{20}$) or only marginally better ($k_{\rm UV}/k_* = 10^{24}$) than constraints from direct gravitational wave detectors. 

Constraints on $n_t$ from direct CMB measurements are obviously dependent on a non-zero detection of $r$, and are unlikely to greatly improve with future measurements given the large foregrounds and the limited lever arm for B-modes. On the other hand, direct measures place useful upper bounds on $n_t$ which are not particularly sensitive to the value of $r$.  In the near future we expect LIGO to improve its bound on gravitational waves and as such on $n_t$. At the same time, the background expansion will also be measured more accurately, with an expected error on the number of massless degrees of freedom of order a percent (see for example \cite{Benson:2014qhw}). Hence, we expect both measures to be equally important. In the long term however, testing the consistency condition would require a more ambitious direct detection experiment with a much higher level of precision. For that reason alone, we anticipate the analysis presented in this paper to be of interest in the future, especially if the value of $r$ is constrained at the percent level. 

\section*{Acknowledgments}
BH was supported by the Astrophysics summer school program. RH would like to thank David Marsh and Kendrick Smith for useful discussions. JM would like to thank Aaron Zimmerman for helpful conversations. PDM would like to thank Dan Grin, Marilena Loverde, Dan Green, Thorsten Battefeld, Marco Peloso, Vitaly Vanchurin, Eiichiro Komatsu and Kipp Cannon for insightful discussions on the gravitational wave energy density. We also like to thank Paul Steinhardt for a very interesting and insightful discussion on inflation.  

\appendix

\section{On the gravitational wave energy density}\label{appendixA}
In this appendix we consider the definition of the energy density of gravitational waves, following the treatment of \cite{Isaacson:1968zza,mtw1973,Komatsu2006,maggiore2007}.  We are ultimately interested in how gravitational waves curve the background spacetime and impact the expansion rate.  We begin by splitting the metric into a smooth background and a perturbation
\begin{equation}
	g_{\mu\nu}=\bar{g}_{\mu\nu}+\delta g_{\mu\nu} \, ,
\end{equation}
where we take the background to be given by the flat Friedmann-Robertson-Walker metric and the perturbation is defined in terms of the gauge invariant tensor perturbation as
\begin{equation}
	\delta g_{ij}=a^2(t)h_{ij} \, ,
\end{equation}
with $h_{ij,j}=0$ and $h_{ii}=0$.  Understanding the impact of the gravitational waves on the evolution of the background requires that we consider only fluctuations for which $k/a(t)\gg H(t)$, modes which are deep inside the horizon.  This might seem to pose a problem for the case we are considering, where there exist gravitational waves with wavelengths comparable to and larger than the horizon size at the epochs of interest.  We will show below, however, that such modes give a negligible contribution to the energy density and can be safely neglected for the cases we consider.

Now let us expand the Einstein equations around the background metric.  There exist two small expansion parameters; the first is the amplitude of the gravitational waves $\delta g\equiv\mathcal{O}(|\delta g_{\mu\nu}|)$, and the second is $k/aH$.  The Ricci tensor can be expanded up to second order in $\delta g_{\mu\nu}$ as
\begin{equation}
	R_{\mu\nu}=\bar{R}_{\mu\nu}+R_{\mu\nu}^{(1)}+R_{\mu\nu}^{(2)}+\cdots \, .
\end{equation}
Now the key point is that $\bar{R}_{\mu\nu}$ is constructed purely from $\bar{g}_{\mu\nu}$ and therefore only contains only low frequency modes (by which we mean variations on scales comparable to $H^{-1}$), while $R_{\mu\nu}^{(1)}$ is linear in $\delta g_{\mu\nu}$ and therefore contains only high frequency modes.  However, $R_{\mu\nu}^{(2)}$ is quadratic in $\delta g_{\mu\nu}$ and therefore contains both high and low frequency modes.  We can therefore split the Einstein equations into low and high frequency parts
\begin{equation}\label{RicciLow}
	\bar{R}_{\mu\nu}=-\left[R_{\mu\nu}^{(2)}\right]^{\mathrm{Low}}+8\pi G\left(T_{\mu\nu}-\frac{1}{2}g_{\mu\nu}T\right)^{\mathrm{Low}} \, ,
\end{equation}
and
\begin{equation}\label{RicciHigh}
	R_{\mu\nu}^{(1)}=-\left[R_{\mu\nu}^{(2)}\right]^{\mathrm{High}}+8\pi G\left(T_{\mu\nu}-\frac{1}{2}g_{\mu\nu}T\right)^{\mathrm{High}} \, .
\end{equation}
Eq.~\eqref{RicciHigh} is a wave equation which governs the propagation of $\delta g_{\mu\nu}$ on the background and is not of immediate interest for us.  Eq.~\eqref{RicciLow} describes how the presence of gravitational waves curves spacetime and allows us to define the energy momentum tensor for gravitational waves.

Since we are considering only perturbations which vary on scales much smaller than the horizon size, we can project onto low frequency modes by performing an average over a region which contains several wavelengths of the gravitational waves (or equivalently over a time in which the gravitational waves undergo many oscillations).  We therefore find
\begin{equation}
	\bar{R}_{\mu\nu}=-\langle R_{\mu\nu}^{(2)}\rangle+8\pi G\langle T_{\mu\nu}-\frac{1}{2}g_{\mu\nu}T\rangle \, ,
\end{equation}
and we can see that gravitational waves have an effective energy momentum tensor which is given by
\begin{equation}
	T_{\mu\nu}^{\mathrm{GW}}=-\frac{1}{8\pi G}\langle R_{\mu\nu}^{(2)}-\frac{1}{2}\bar{g}_{\mu\nu}R^{(2)}\rangle+\mathcal{O}(\delta g^3) \, .
\end{equation}
This quantity can be computed explicitly in the transverse traceless gauge as
\begin{equation}
	T_{\mu\nu}^{\mathrm{GW}}=\frac{1}{32\pi G}\bar{g}^{\alpha\rho}\bar{g}^{\beta\sigma}\langle \delta g_{\alpha\beta |\mu}\delta g_{\rho\sigma |\nu}\rangle + \mathcal{O}(\delta g^3) \, ,
\end{equation}
where a vertical bar indicates a covariant derivative with respect to the background metric.  The energy density of gravitational waves is then given by
\begin{align}
	\rho_{\mathrm{GW}}&=T_{00}^{\mathrm{GW}} \nonumber \\
	&=\frac{1}{32 \pi Ga^4}\delta^{ik}\delta^{j\ell}\langle (\partial_0-2H)\delta g_{ij}(\partial_0-2H)\delta g_{k\ell}\rangle  \nonumber \\ 
	&+ \mathcal{O}(\delta g^3) \nonumber \\
	&=\frac{1}{32\pi G}\delta^{ik}\delta^{j\ell}\langle\dot{h}_{ij}\dot{h}_{k\ell}\rangle+\mathcal{O}(\delta g^3) \, .
\end{align}
With the power spectrum $\Delta_h^2(\eta,k)$ defined as
\begin{equation}
	\langle h_{ij}(\eta,\mathbf{x}) h^{ij}(\eta,\mathbf{x}) \rangle\equiv\int d\log k\Delta_h^2(\eta,k) \, ,
\end{equation}
and the transfer function $\mathcal{T}(\eta,k)$ defined as
\begin{equation}
	\Delta_h^2(k,\eta)=P_t(k)\left[\mathcal{T}(\eta,k)\right]^2\, ,
\end{equation}
the energy density of gravitational waves can be written as
\begin{equation}\label{GWenergy}
	\rho_{\mathrm{GW}}=\frac{1}{32\pi G a^2}\int d\log k P_t(k)\left[\mathcal{T}'(\eta,k)\right]^2 \, ,
\end{equation}
where a prime refers to a derivative with respect to conformal time $\eta$.

Constraints on direct detection of gravitational waves tend to be quoted in terms of the normalized energy density per logarithmic scale
\begin{equation}
	\Omega_{GW}(k)\equiv\frac{1}{\rho_{\mathrm{crit},0}}\frac{d\rho_{\mathrm{GW}}}{d\log k}=\frac{P_t(k)}{12H_0^2a^2}\left[\mathcal{T}'(\eta_0,k)\right]^2 \, ,
\end{equation}
where we have used the fact that the critical energy density today is given by $\rho_{\mathrm{crit},0}=3H_0^2/8\pi G$.

Now let us briefly comment on the limits of the integral over wavenumber appearing in Eq.~\eqref{GWenergy}.  In order to be consistent, we should fix the lower limit to be at a scale for which $k_{\mathrm{IR}}/a(t)\gg H(t)$ since we restricted ourselves in this derivation to modes which were deep inside the horizon.  On the other hand, the derivative of the transfer function scales as $k^2$ in the limit that $k\rightarrow 0$ \cite{Komatsu2006}.  This behavior ensures that the integral will converge in the limit $k_{\mathrm{IR}}\rightarrow 0$ for $n_t>-4$. Furthermore the long wavelength modes (those for which $k/a\lesssim \frac{1}{10}H$) will make a negligible contribution to the total energy density for $n_t\gtrsim-2$.  Since we are mostly interested in cases where $n_t>0$, we can safely take $k_{\mathrm{IR}}=0$ without introducing a large error.  The upper limit on the integral generally depends on the mechanism by which the gravitational waves were produced in the early universe.  We consider a few cases in the main text and discuss how our constraints depend on this choice.

\bibliography{GW_paper} 

\begin{thebibliography}{53}
\expandafter\ifx\csname natexlab\endcsname\relax\def\natexlab#1{#1}\fi
\expandafter\ifx\csname bibnamefont\endcsname\relax
  \def\bibnamefont#1{#1}\fi
\expandafter\ifx\csname bibfnamefont\endcsname\relax
  \def\bibfnamefont#1{#1}\fi
\expandafter\ifx\csname citenamefont\endcsname\relax
  \def\citenamefont#1{#1}\fi
\expandafter\ifx\csname url\endcsname\relax
  \def\url#1{\texttt{#1}}\fi
\expandafter\ifx\csname urlprefix\endcsname\relax\def\urlprefix{URL }\fi
\providecommand{\bibinfo}[2]{#2}
\providecommand{\eprint}[2][]{\url{#2}}

\bibitem[{\citenamefont{{Ade} et~al.}(2014)\citenamefont{{Ade}, {Aikin},
  {Barkats}, {Benton}, {Bischoff}, {Bock}, {Brevik}, {Buder}, {Bullock},
  {Dowell} et~al.}}]{BICEP2}
\bibinfo{author}{\bibfnamefont{P.~A.~R.} \bibnamefont{{Ade}}},
  \bibinfo{author}{\bibfnamefont{R.~W.} \bibnamefont{{Aikin}}},
  \bibinfo{author}{\bibfnamefont{D.}~\bibnamefont{{Barkats}}},
  \bibinfo{author}{\bibfnamefont{S.~J.} \bibnamefont{{Benton}}},
  \bibinfo{author}{\bibfnamefont{C.~A.} \bibnamefont{{Bischoff}}},
  \bibinfo{author}{\bibfnamefont{J.~J.} \bibnamefont{{Bock}}},
  \bibinfo{author}{\bibfnamefont{J.~A.} \bibnamefont{{Brevik}}},
  \bibinfo{author}{\bibfnamefont{I.}~\bibnamefont{{Buder}}},
  \bibinfo{author}{\bibfnamefont{E.}~\bibnamefont{{Bullock}}},
  \bibinfo{author}{\bibfnamefont{C.~D.} \bibnamefont{{Dowell}}},
  \bibnamefont{et~al.}, \bibinfo{journal}{Physical Review Letters}
  \textbf{\bibinfo{volume}{112}}, \bibinfo{eid}{241101} (\bibinfo{year}{2014}),
  \eprint{1403.3985}.

\bibitem[{\citenamefont{{BICEP2/Keck} et~al.}(2015)\citenamefont{{BICEP2/Keck},
  {Planck Collaborations}, {:}, {Ade}, {Aghanim}, {Ahmed}, {Aikin},
  {Alexander}, {Arnaud}, {Aumont} et~al.}}]{BICEPKECK2015}
\bibinfo{author}{\bibnamefont{{BICEP2/Keck}}},
  \bibinfo{author}{\bibnamefont{{Planck Collaborations}}},
  \bibinfo{author}{\bibnamefont{{:}}}, \bibinfo{author}{\bibfnamefont{P.~A.~R.}
  \bibnamefont{{Ade}}},
  \bibinfo{author}{\bibfnamefont{N.}~\bibnamefont{{Aghanim}}},
  \bibinfo{author}{\bibfnamefont{Z.}~\bibnamefont{{Ahmed}}},
  \bibinfo{author}{\bibfnamefont{R.~W.} \bibnamefont{{Aikin}}},
  \bibinfo{author}{\bibfnamefont{K.~D.} \bibnamefont{{Alexander}}},
  \bibinfo{author}{\bibfnamefont{M.}~\bibnamefont{{Arnaud}}},
  \bibinfo{author}{\bibfnamefont{J.}~\bibnamefont{{Aumont}}},
  \bibnamefont{et~al.}, \bibinfo{journal}{ArXiv e-prints}
  (\bibinfo{year}{2015}), \eprint{1502.00612}.

\bibitem[{\citenamefont{{Liddle} and {Lyth}}(1993)}]{Liddle1993}
\bibinfo{author}{\bibfnamefont{A.~R.} \bibnamefont{{Liddle}}} \bibnamefont{and}
  \bibinfo{author}{\bibfnamefont{D.~H.} \bibnamefont{{Lyth}}},
  \bibinfo{journal}{\mnras} \textbf{\bibinfo{volume}{265}},
  \bibinfo{pages}{379} (\bibinfo{year}{1993}), \eprint{astro-ph/9304017}.

\bibitem[{\citenamefont{{Liddle} and {Lyth}}(2000)}]{liddle_lyth}
\bibinfo{author}{\bibfnamefont{A.~R.} \bibnamefont{{Liddle}}} \bibnamefont{and}
  \bibinfo{author}{\bibfnamefont{D.~H.} \bibnamefont{{Lyth}}},
  \emph{\bibinfo{title}{{Cosmological Inflation and Large-Scale Structure}}}
  (\bibinfo{year}{2000}).

\bibitem[{\citenamefont{{Liddle} and {Lyth}}(1992)}]{1992PhLB..291..391L}
\bibinfo{author}{\bibfnamefont{A.~R.} \bibnamefont{{Liddle}}} \bibnamefont{and}
  \bibinfo{author}{\bibfnamefont{D.~H.} \bibnamefont{{Lyth}}},
  \bibinfo{journal}{Physics Letters B} \textbf{\bibinfo{volume}{291}},
  \bibinfo{pages}{391} (\bibinfo{year}{1992}), \eprint{astro-ph/9208007}.

\bibitem[{\citenamefont{{Copeland} et~al.}(1993)\citenamefont{{Copeland},
  {Kolb}, {Liddle}, and {Lidsey}}}]{1993PhRvL..71..219C}
\bibinfo{author}{\bibfnamefont{E.~J.} \bibnamefont{{Copeland}}},
  \bibinfo{author}{\bibfnamefont{E.~W.} \bibnamefont{{Kolb}}},
  \bibinfo{author}{\bibfnamefont{A.~R.} \bibnamefont{{Liddle}}},
  \bibnamefont{and} \bibinfo{author}{\bibfnamefont{J.~E.}
  \bibnamefont{{Lidsey}}}, \bibinfo{journal}{Physical Review Letters}
  \textbf{\bibinfo{volume}{71}}, \bibinfo{pages}{219} (\bibinfo{year}{1993}),
  \eprint{hep-ph/9304228}.

\bibitem[{\citenamefont{{Price} et~al.}(2014)\citenamefont{{Price}, {Peiris},
  {Frazer}, and {Easther}}}]{2014arXiv1409.2498P}
\bibinfo{author}{\bibfnamefont{L.~C.} \bibnamefont{{Price}}},
  \bibinfo{author}{\bibfnamefont{H.~V.} \bibnamefont{{Peiris}}},
  \bibinfo{author}{\bibfnamefont{J.}~\bibnamefont{{Frazer}}}, \bibnamefont{and}
  \bibinfo{author}{\bibfnamefont{R.}~\bibnamefont{{Easther}}},
  \bibinfo{journal}{ArXiv e-prints}  (\bibinfo{year}{2014}),
  \eprint{1409.2498}.

\bibitem[{\citenamefont{{Baumann} et~al.}(2014)\citenamefont{{Baumann},
  {Green}, and {Porto}}}]{2014arXiv1407.2621B}
\bibinfo{author}{\bibfnamefont{D.}~\bibnamefont{{Baumann}}},
  \bibinfo{author}{\bibfnamefont{D.}~\bibnamefont{{Green}}}, \bibnamefont{and}
  \bibinfo{author}{\bibfnamefont{R.~A.} \bibnamefont{{Porto}}},
  \bibinfo{journal}{ArXiv e-prints}  (\bibinfo{year}{2014}),
  \eprint{1407.2621}.

\bibitem[{\citenamefont{{Palma} and {Soto}}(2014)}]{palma/soto:2014}
\bibinfo{author}{\bibfnamefont{G.~A.} \bibnamefont{{Palma}}} \bibnamefont{and}
  \bibinfo{author}{\bibfnamefont{A.}~\bibnamefont{{Soto}}},
  \bibinfo{journal}{ArXiv e-prints}  (\bibinfo{year}{2014}),
  \eprint{1412.0033}.

\bibitem[{\citenamefont{{Lehners}}(2008)}]{2008PhR...465..223L}
\bibinfo{author}{\bibfnamefont{J.-L.} \bibnamefont{{Lehners}}},
  \bibinfo{journal}{\physrep} \textbf{\bibinfo{volume}{465}},
  \bibinfo{pages}{223} (\bibinfo{year}{2008}), \eprint{0806.1245}.

\bibitem[{\citenamefont{{Boyle} et~al.}(2004)\citenamefont{{Boyle},
  {Steinhardt}, and {Turok}}}]{2004PhRvD..69l7302B}
\bibinfo{author}{\bibfnamefont{L.~A.} \bibnamefont{{Boyle}}},
  \bibinfo{author}{\bibfnamefont{P.~J.} \bibnamefont{{Steinhardt}}},
  \bibnamefont{and} \bibinfo{author}{\bibfnamefont{N.}~\bibnamefont{{Turok}}},
  \bibinfo{journal}{\prd} \textbf{\bibinfo{volume}{69}}, \bibinfo{eid}{127302}
  (\bibinfo{year}{2004}), \eprint{hep-th/0307170}.

\bibitem[{\citenamefont{{Brandenberger}
  et~al.}(2007)\citenamefont{{Brandenberger}, {Nayeri}, {Patil}, and
  {Vafa}}}]{2007PhRvL..98w1302B}
\bibinfo{author}{\bibfnamefont{R.~H.} \bibnamefont{{Brandenberger}}},
  \bibinfo{author}{\bibfnamefont{A.}~\bibnamefont{{Nayeri}}},
  \bibinfo{author}{\bibfnamefont{S.~P.} \bibnamefont{{Patil}}},
  \bibnamefont{and} \bibinfo{author}{\bibfnamefont{C.}~\bibnamefont{{Vafa}}},
  \bibinfo{journal}{Physical Review Letters} \textbf{\bibinfo{volume}{98}},
  \bibinfo{eid}{231302} (\bibinfo{year}{2007}), \eprint{hep-th/0604126}.

\bibitem[{\citenamefont{{Senatore} et~al.}(2014)\citenamefont{{Senatore},
  {Silverstein}, and {Zaldarriaga}}}]{2014JCAP...08..016S}
\bibinfo{author}{\bibfnamefont{L.}~\bibnamefont{{Senatore}}},
  \bibinfo{author}{\bibfnamefont{E.}~\bibnamefont{{Silverstein}}},
  \bibnamefont{and}
  \bibinfo{author}{\bibfnamefont{M.}~\bibnamefont{{Zaldarriaga}}},
  \bibinfo{journal}{\jcap} \textbf{\bibinfo{volume}{8}}, \bibinfo{eid}{016}
  (\bibinfo{year}{2014}), \eprint{1109.0542}.

\bibitem[{\citenamefont{{Mirbabayi} et~al.}(2014)\citenamefont{{Mirbabayi},
  {Senatore}, {Silverstein}, and {Zaldarriaga}}}]{2014arXiv1412.0665M}
\bibinfo{author}{\bibfnamefont{M.}~\bibnamefont{{Mirbabayi}}},
  \bibinfo{author}{\bibfnamefont{L.}~\bibnamefont{{Senatore}}},
  \bibinfo{author}{\bibfnamefont{E.}~\bibnamefont{{Silverstein}}},
  \bibnamefont{and}
  \bibinfo{author}{\bibfnamefont{M.}~\bibnamefont{{Zaldarriaga}}},
  \bibinfo{journal}{ArXiv e-prints}  (\bibinfo{year}{2014}),
  \eprint{1412.0665}.

\bibitem[{\citenamefont{{Barnaby} et~al.}(2012)\citenamefont{{Barnaby},
  {Pajer}, and {Peloso}}}]{2012PhRvD..85b3525B}
\bibinfo{author}{\bibfnamefont{N.}~\bibnamefont{{Barnaby}}},
  \bibinfo{author}{\bibfnamefont{E.}~\bibnamefont{{Pajer}}}, \bibnamefont{and}
  \bibinfo{author}{\bibfnamefont{M.}~\bibnamefont{{Peloso}}},
  \bibinfo{journal}{\prd} \textbf{\bibinfo{volume}{85}}, \bibinfo{eid}{023525}
  (\bibinfo{year}{2012}), \eprint{1110.3327}.

\bibitem[{\citenamefont{{Abbott} et~al.}(2009)\citenamefont{{Abbott}, {Abbott},
  {Adhikari}, {Ajith}, {Allen}, {Allen}, {Amin}, {Anderson}, {Anderson},
  {Arain} et~al.}}]{abbott/etal:2007}
\bibinfo{author}{\bibfnamefont{B.~P.} \bibnamefont{{Abbott}}},
  \bibinfo{author}{\bibfnamefont{R.}~\bibnamefont{{Abbott}}},
  \bibinfo{author}{\bibfnamefont{R.}~\bibnamefont{{Adhikari}}},
  \bibinfo{author}{\bibfnamefont{P.}~\bibnamefont{{Ajith}}},
  \bibinfo{author}{\bibfnamefont{B.}~\bibnamefont{{Allen}}},
  \bibinfo{author}{\bibfnamefont{G.}~\bibnamefont{{Allen}}},
  \bibinfo{author}{\bibfnamefont{R.~S.} \bibnamefont{{Amin}}},
  \bibinfo{author}{\bibfnamefont{S.~B.} \bibnamefont{{Anderson}}},
  \bibinfo{author}{\bibfnamefont{W.~G.} \bibnamefont{{Anderson}}},
  \bibinfo{author}{\bibfnamefont{M.~A.} \bibnamefont{{Arain}}},
  \bibnamefont{et~al.}, \bibinfo{journal}{Reports on Progress in Physics}
  \textbf{\bibinfo{volume}{72}}, \bibinfo{eid}{076901} (\bibinfo{year}{2009}),
  \eprint{0711.3041}.

\bibitem[{\citenamefont{{Manchester} et~al.}(2013)\citenamefont{{Manchester},
  {Hobbs}, {Bailes}, {Coles}, {van Straten}, {Keith}, {Shannon}, {Bhat},
  {Brown}, {Burke-Spolaor} et~al.}}]{manchester/etal:2013}
\bibinfo{author}{\bibfnamefont{R.~N.} \bibnamefont{{Manchester}}},
  \bibinfo{author}{\bibfnamefont{G.}~\bibnamefont{{Hobbs}}},
  \bibinfo{author}{\bibfnamefont{M.}~\bibnamefont{{Bailes}}},
  \bibinfo{author}{\bibfnamefont{W.~A.} \bibnamefont{{Coles}}},
  \bibinfo{author}{\bibfnamefont{W.}~\bibnamefont{{van Straten}}},
  \bibinfo{author}{\bibfnamefont{M.~J.} \bibnamefont{{Keith}}},
  \bibinfo{author}{\bibfnamefont{R.~M.} \bibnamefont{{Shannon}}},
  \bibinfo{author}{\bibfnamefont{N.~D.~R.} \bibnamefont{{Bhat}}},
  \bibinfo{author}{\bibfnamefont{A.}~\bibnamefont{{Brown}}},
  \bibinfo{author}{\bibfnamefont{S.~G.} \bibnamefont{{Burke-Spolaor}}},
  \bibnamefont{et~al.}, \bibinfo{journal}{\pasp} \textbf{\bibinfo{volume}{30}},
  \bibinfo{eid}{e017} (\bibinfo{year}{2013}), \eprint{1210.6130}.

\bibitem[{\citenamefont{Smith et~al.}(2006)\citenamefont{Smith, Peiris, and
  Cooray}}]{smith/peiris/cooray:2006}
\bibinfo{author}{\bibfnamefont{T.~L.} \bibnamefont{Smith}},
  \bibinfo{author}{\bibfnamefont{H.~V.} \bibnamefont{Peiris}},
  \bibnamefont{and} \bibinfo{author}{\bibfnamefont{A.}~\bibnamefont{Cooray}},
  \bibinfo{journal}{Phys.Rev.} \textbf{\bibinfo{volume}{D73}},
  \bibinfo{pages}{123503} (\bibinfo{year}{2006}), \eprint{astro-ph/0602137}.

\bibitem[{\citenamefont{{Cort{\^e}s} et~al.}(2014)\citenamefont{{Cort{\^e}s},
  {Liddle}, and {Parkinson}}}]{2014arXiv1409.6530C}
\bibinfo{author}{\bibfnamefont{M.}~\bibnamefont{{Cort{\^e}s}}},
  \bibinfo{author}{\bibfnamefont{A.~R.} \bibnamefont{{Liddle}}},
  \bibnamefont{and}
  \bibinfo{author}{\bibfnamefont{D.}~\bibnamefont{{Parkinson}}},
  \bibinfo{journal}{ArXiv e-prints}  (\bibinfo{year}{2014}),
  \eprint{1409.6530}.

\bibitem[{\citenamefont{{Caligiuri} and {Kosowsky}}(2014)}]{Caligiuri2014}
\bibinfo{author}{\bibfnamefont{J.}~\bibnamefont{{Caligiuri}}} \bibnamefont{and}
  \bibinfo{author}{\bibfnamefont{A.}~\bibnamefont{{Kosowsky}}},
  \bibinfo{journal}{Physical Review Letters} \textbf{\bibinfo{volume}{112}},
  \bibinfo{eid}{191302} (\bibinfo{year}{2014}), \eprint{1403.5324}.

\bibitem[{\citenamefont{{Boyle} and {Buonanno}}(2008)}]{Boyle2007}
\bibinfo{author}{\bibfnamefont{L.~A.} \bibnamefont{{Boyle}}} \bibnamefont{and}
  \bibinfo{author}{\bibfnamefont{A.}~\bibnamefont{{Buonanno}}},
  \bibinfo{journal}{\prd} \textbf{\bibinfo{volume}{78}}, \bibinfo{eid}{043531}
  (\bibinfo{year}{2008}), \eprint{0708.2279}.

\bibitem[{\citenamefont{Misner et~al.}(1973)\citenamefont{Misner, Thorne, and
  Wheeler}}]{mtw1973}
\bibinfo{author}{\bibfnamefont{C.~W.} \bibnamefont{Misner}},
  \bibinfo{author}{\bibfnamefont{K.~S.} \bibnamefont{Thorne}},
  \bibnamefont{and} \bibinfo{author}{\bibfnamefont{J.~A.}
  \bibnamefont{Wheeler}}, \emph{\bibinfo{title}{Gravitation}}
  (\bibinfo{publisher}{W. H. Freeman San Francisco}, \bibinfo{year}{1973}),
  ISBN \bibinfo{isbn}{0716703343 0716703440}.

\bibitem[{\citenamefont{Krauss and White}(1992)}]{krauss/white:1992}
\bibinfo{author}{\bibfnamefont{L.~M.} \bibnamefont{Krauss}} \bibnamefont{and}
  \bibinfo{author}{\bibfnamefont{M.}~\bibnamefont{White}},
  \bibinfo{journal}{Phys. Rev. Lett.} \textbf{\bibinfo{volume}{69}},
  \bibinfo{pages}{869} (\bibinfo{year}{1992}),
  \urlprefix\url{http://link.aps.org/doi/10.1103/PhysRevLett.69.869}.

\bibitem[{\citenamefont{{Kuroyanagi} et~al.}(2014)\citenamefont{{Kuroyanagi},
  {Takahashi}, and {Yokoyama}}}]{2014arXiv1407.4785K}
\bibinfo{author}{\bibfnamefont{S.}~\bibnamefont{{Kuroyanagi}}},
  \bibinfo{author}{\bibfnamefont{T.}~\bibnamefont{{Takahashi}}},
  \bibnamefont{and}
  \bibinfo{author}{\bibfnamefont{S.}~\bibnamefont{{Yokoyama}}},
  \bibinfo{journal}{ArXiv e-prints}  (\bibinfo{year}{2014}),
  \eprint{1407.4785}.

\bibitem[{\citenamefont{{Liddle} et~al.}(1994)\citenamefont{{Liddle},
  {Parsons}, and {Barrow}}}]{Liddle1994}
\bibinfo{author}{\bibfnamefont{A.~R.} \bibnamefont{{Liddle}}},
  \bibinfo{author}{\bibfnamefont{P.}~\bibnamefont{{Parsons}}},
  \bibnamefont{and} \bibinfo{author}{\bibfnamefont{J.~D.}
  \bibnamefont{{Barrow}}}, \bibinfo{journal}{\prd}
  \textbf{\bibinfo{volume}{50}}, \bibinfo{pages}{7222} (\bibinfo{year}{1994}),
  \eprint{astro-ph/9408015}.

\bibitem[{\citenamefont{{Boyle} et~al.}(2014)\citenamefont{{Boyle}, {Smith},
  {Dvorkin}, and {Turok}}}]{boyle/etal:2014}
\bibinfo{author}{\bibfnamefont{L.}~\bibnamefont{{Boyle}}},
  \bibinfo{author}{\bibfnamefont{K.~M.} \bibnamefont{{Smith}}},
  \bibinfo{author}{\bibfnamefont{C.}~\bibnamefont{{Dvorkin}}},
  \bibnamefont{and} \bibinfo{author}{\bibfnamefont{N.}~\bibnamefont{{Turok}}},
  \bibinfo{journal}{ArXiv e-prints}  (\bibinfo{year}{2014}),
  \eprint{1408.3129}.

\bibitem[{\citenamefont{{Page} et~al.}(2007)\citenamefont{{Page}, {Hinshaw},
  {Komatsu}, {Nolta}, {Spergel}, {Bennett}, {Barnes}, {Bean}, {Dor{\'e}},
  {Dunkley} et~al.}}]{WMAP}
\bibinfo{author}{\bibfnamefont{L.}~\bibnamefont{{Page}}},
  \bibinfo{author}{\bibfnamefont{G.}~\bibnamefont{{Hinshaw}}},
  \bibinfo{author}{\bibfnamefont{E.}~\bibnamefont{{Komatsu}}},
  \bibinfo{author}{\bibfnamefont{M.~R.} \bibnamefont{{Nolta}}},
  \bibinfo{author}{\bibfnamefont{D.~N.} \bibnamefont{{Spergel}}},
  \bibinfo{author}{\bibfnamefont{C.~L.} \bibnamefont{{Bennett}}},
  \bibinfo{author}{\bibfnamefont{C.}~\bibnamefont{{Barnes}}},
  \bibinfo{author}{\bibfnamefont{R.}~\bibnamefont{{Bean}}},
  \bibinfo{author}{\bibfnamefont{O.}~\bibnamefont{{Dor{\'e}}}},
  \bibinfo{author}{\bibfnamefont{J.}~\bibnamefont{{Dunkley}}},
  \bibnamefont{et~al.}, \bibinfo{journal}{\apjs}
  \textbf{\bibinfo{volume}{170}}, \bibinfo{pages}{335} (\bibinfo{year}{2007}),
  \eprint{astro-ph/0603450}.

\bibitem[{\citenamefont{{Planck Collaboration}
  et~al.}(2013)\citenamefont{{Planck Collaboration}, {Ade}, {Aghanim},
  {Armitage-Caplan}, {Arnaud}, {Ashdown}, {Atrio-Barandela}, {Aumont},
  {Baccigalupi}, {Banday} et~al.}}]{Planck}
\bibinfo{author}{\bibnamefont{{Planck Collaboration}}},
  \bibinfo{author}{\bibfnamefont{P.~A.~R.} \bibnamefont{{Ade}}},
  \bibinfo{author}{\bibfnamefont{N.}~\bibnamefont{{Aghanim}}},
  \bibinfo{author}{\bibfnamefont{C.}~\bibnamefont{{Armitage-Caplan}}},
  \bibinfo{author}{\bibfnamefont{M.}~\bibnamefont{{Arnaud}}},
  \bibinfo{author}{\bibfnamefont{M.}~\bibnamefont{{Ashdown}}},
  \bibinfo{author}{\bibfnamefont{F.}~\bibnamefont{{Atrio-Barandela}}},
  \bibinfo{author}{\bibfnamefont{J.}~\bibnamefont{{Aumont}}},
  \bibinfo{author}{\bibfnamefont{C.}~\bibnamefont{{Baccigalupi}}},
  \bibinfo{author}{\bibfnamefont{A.~J.} \bibnamefont{{Banday}}},
  \bibnamefont{et~al.}, \bibinfo{journal}{ArXiv e-prints}
  (\bibinfo{year}{2013}), \eprint{1303.5076}.

\bibitem[{\citenamefont{{Scoville} et~al.}(2007)\citenamefont{{Scoville},
  {Abraham}, {Aussel}, {Barnes}, {Benson}, {Blain}, {Calzetti}, {Comastri},
  {Capak}, {Carilli} et~al.}}]{HST}
\bibinfo{author}{\bibfnamefont{N.}~\bibnamefont{{Scoville}}},
  \bibinfo{author}{\bibfnamefont{R.~G.} \bibnamefont{{Abraham}}},
  \bibinfo{author}{\bibfnamefont{H.}~\bibnamefont{{Aussel}}},
  \bibinfo{author}{\bibfnamefont{J.~E.} \bibnamefont{{Barnes}}},
  \bibinfo{author}{\bibfnamefont{A.}~\bibnamefont{{Benson}}},
  \bibinfo{author}{\bibfnamefont{A.~W.} \bibnamefont{{Blain}}},
  \bibinfo{author}{\bibfnamefont{D.}~\bibnamefont{{Calzetti}}},
  \bibinfo{author}{\bibfnamefont{A.}~\bibnamefont{{Comastri}}},
  \bibinfo{author}{\bibfnamefont{P.}~\bibnamefont{{Capak}}},
  \bibinfo{author}{\bibfnamefont{C.}~\bibnamefont{{Carilli}}},
  \bibnamefont{et~al.}, \bibinfo{journal}{\apjs}
  \textbf{\bibinfo{volume}{172}}, \bibinfo{pages}{38} (\bibinfo{year}{2007}),
  \eprint{astro-ph/0612306}.

\bibitem[{\citenamefont{{York} et~al.}(2000)\citenamefont{{York}, {Adelman},
  {Anderson}, {Anderson}, {Annis}, {Bahcall}, {Bakken}, {Barkhouser},
  {Bastian}, {Berman} et~al.}}]{BAO}
\bibinfo{author}{\bibfnamefont{D.~G.} \bibnamefont{{York}}},
  \bibinfo{author}{\bibfnamefont{J.}~\bibnamefont{{Adelman}}},
  \bibinfo{author}{\bibfnamefont{J.~E.} \bibnamefont{{Anderson}},
  \bibfnamefont{Jr.}}, \bibinfo{author}{\bibfnamefont{S.~F.}
  \bibnamefont{{Anderson}}},
  \bibinfo{author}{\bibfnamefont{J.}~\bibnamefont{{Annis}}},
  \bibinfo{author}{\bibfnamefont{N.~A.} \bibnamefont{{Bahcall}}},
  \bibinfo{author}{\bibfnamefont{J.~A.} \bibnamefont{{Bakken}}},
  \bibinfo{author}{\bibfnamefont{R.}~\bibnamefont{{Barkhouser}}},
  \bibinfo{author}{\bibfnamefont{S.}~\bibnamefont{{Bastian}}},
  \bibinfo{author}{\bibfnamefont{E.}~\bibnamefont{{Berman}}},
  \bibnamefont{et~al.}, \bibinfo{journal}{\aj} \textbf{\bibinfo{volume}{120}},
  \bibinfo{pages}{1579} (\bibinfo{year}{2000}), \eprint{astro-ph/0006396}.

\bibitem[{\citenamefont{{Lewis} and {Bridle}}(2002)}]{cosmomc}
\bibinfo{author}{\bibfnamefont{A.}~\bibnamefont{{Lewis}}} \bibnamefont{and}
  \bibinfo{author}{\bibfnamefont{S.}~\bibnamefont{{Bridle}}},
  \bibinfo{journal}{\prd} \textbf{\bibinfo{volume}{66}}, \bibinfo{eid}{103511}
  (\bibinfo{year}{2002}), \eprint{arXiv:astro-ph/0205436}.

\bibitem[{\citenamefont{Lewis et~al.}(2000)\citenamefont{Lewis, Challinor, and
  Lasenby}}]{Lewis:1999bs}
\bibinfo{author}{\bibfnamefont{A.}~\bibnamefont{Lewis}},
  \bibinfo{author}{\bibfnamefont{A.}~\bibnamefont{Challinor}},
  \bibnamefont{and} \bibinfo{author}{\bibfnamefont{A.}~\bibnamefont{Lasenby}},
  \bibinfo{journal}{Astrophys. J.} \textbf{\bibinfo{volume}{538}},
  \bibinfo{pages}{473} (\bibinfo{year}{2000}), \eprint{astro-ph/9911177}.

\bibitem[{\citenamefont{{Moore} et~al.}(2014)\citenamefont{{Moore}, {Cole}, and
  {Berry}}}]{Moore2014}
\bibinfo{author}{\bibfnamefont{C.~J.} \bibnamefont{{Moore}}},
  \bibinfo{author}{\bibfnamefont{R.~H.} \bibnamefont{{Cole}}},
  \bibnamefont{and} \bibinfo{author}{\bibfnamefont{C.~P.~L.}
  \bibnamefont{{Berry}}}, \bibinfo{journal}{ArXiv e-prints}
  (\bibinfo{year}{2014}), \eprint{1408.0740}.

\bibitem[{\citenamefont{{The LIGO Scientific Collaboration}
  et~al.}(2014)\citenamefont{{The LIGO Scientific Collaboration}, {the Virgo
  Collaboration}, {Aasi}, {Abbott}, {Abbott}, {Abbott}, {Abernathy}, {Accadia},
  {Acernese}, {Ackley} et~al.}}]{LIGO}
\bibinfo{author}{\bibnamefont{{The LIGO Scientific Collaboration}}},
  \bibinfo{author}{\bibnamefont{{the Virgo Collaboration}}},
  \bibinfo{author}{\bibfnamefont{J.}~\bibnamefont{{Aasi}}},
  \bibinfo{author}{\bibfnamefont{B.~P.} \bibnamefont{{Abbott}}},
  \bibinfo{author}{\bibfnamefont{R.}~\bibnamefont{{Abbott}}},
  \bibinfo{author}{\bibfnamefont{T.}~\bibnamefont{{Abbott}}},
  \bibinfo{author}{\bibfnamefont{M.~R.} \bibnamefont{{Abernathy}}},
  \bibinfo{author}{\bibfnamefont{T.}~\bibnamefont{{Accadia}}},
  \bibinfo{author}{\bibfnamefont{F.}~\bibnamefont{{Acernese}}},
  \bibinfo{author}{\bibfnamefont{K.}~\bibnamefont{{Ackley}}},
  \bibnamefont{et~al.}, \bibinfo{journal}{ArXiv e-prints}
  (\bibinfo{year}{2014}), \eprint{1406.4556}.

\bibitem[{\citenamefont{{Shannon} et~al.}(2013)\citenamefont{{Shannon}, {Ravi},
  {Coles}, {Hobbs}, {Keith}, {Manchester}, {Wyithe}, {Bailes}, {Bhat},
  {Burke-Spolaor} et~al.}}]{PPTA}
\bibinfo{author}{\bibfnamefont{R.~M.} \bibnamefont{{Shannon}}},
  \bibinfo{author}{\bibfnamefont{V.}~\bibnamefont{{Ravi}}},
  \bibinfo{author}{\bibfnamefont{W.~A.} \bibnamefont{{Coles}}},
  \bibinfo{author}{\bibfnamefont{G.}~\bibnamefont{{Hobbs}}},
  \bibinfo{author}{\bibfnamefont{M.~J.} \bibnamefont{{Keith}}},
  \bibinfo{author}{\bibfnamefont{R.~N.} \bibnamefont{{Manchester}}},
  \bibinfo{author}{\bibfnamefont{J.~S.~B.} \bibnamefont{{Wyithe}}},
  \bibinfo{author}{\bibfnamefont{M.}~\bibnamefont{{Bailes}}},
  \bibinfo{author}{\bibfnamefont{N.~D.~R.} \bibnamefont{{Bhat}}},
  \bibinfo{author}{\bibfnamefont{S.}~\bibnamefont{{Burke-Spolaor}}},
  \bibnamefont{et~al.}, \bibinfo{journal}{Science}
  \textbf{\bibinfo{volume}{342}}, \bibinfo{pages}{334} (\bibinfo{year}{2013}),
  \eprint{1310.4569}.

\bibitem[{\citenamefont{{Sazhin} et~al.}(2011)\citenamefont{{Sazhin},
  {Sazhina}, and {Chadayammuri}}}]{Sazhin2011}
\bibinfo{author}{\bibfnamefont{M.~V.} \bibnamefont{{Sazhin}}},
  \bibinfo{author}{\bibfnamefont{O.~S.} \bibnamefont{{Sazhina}}},
  \bibnamefont{and}
  \bibinfo{author}{\bibfnamefont{U.}~\bibnamefont{{Chadayammuri}}},
  \bibinfo{journal}{ArXiv e-prints}  (\bibinfo{year}{2011}),
  \eprint{1109.2258}.

\bibitem[{\citenamefont{{Dodelson}}(2014)}]{Dodelson}
\bibinfo{author}{\bibfnamefont{S.}~\bibnamefont{{Dodelson}}},
  \bibinfo{journal}{Physical Review Letters} \textbf{\bibinfo{volume}{112}},
  \bibinfo{eid}{191301} (\bibinfo{year}{2014}), \eprint{1403.6310}.

\bibitem[{\citenamefont{{Smith} et~al.}(2006)\citenamefont{{Smith},
  {Pierpaoli}, and {Kamionkowski}}}]{NeffSmith2006}
\bibinfo{author}{\bibfnamefont{T.~L.} \bibnamefont{{Smith}}},
  \bibinfo{author}{\bibfnamefont{E.}~\bibnamefont{{Pierpaoli}}},
  \bibnamefont{and}
  \bibinfo{author}{\bibfnamefont{M.}~\bibnamefont{{Kamionkowski}}},
  \bibinfo{journal}{Physical Review Letters} \textbf{\bibinfo{volume}{97}},
  \bibinfo{eid}{021301} (\bibinfo{year}{2006}), \eprint{astro-ph/0603144}.

\bibitem[{\citenamefont{{Bashinsky} and {Seljak}}(2004)}]{2004PhRvD..69h3002B}
\bibinfo{author}{\bibfnamefont{S.}~\bibnamefont{{Bashinsky}}} \bibnamefont{and}
  \bibinfo{author}{\bibfnamefont{U.}~\bibnamefont{{Seljak}}},
  \bibinfo{journal}{\prd} \textbf{\bibinfo{volume}{69}}, \bibinfo{eid}{083002}
  (\bibinfo{year}{2004}), \eprint{astro-ph/0310198}.

\bibitem[{\citenamefont{{Hou} et~al.}(2013)\citenamefont{{Hou}, {Keisler},
  {Knox}, {Millea}, and {Reichardt}}}]{2013PhRvD..87h3008H}
\bibinfo{author}{\bibfnamefont{Z.}~\bibnamefont{{Hou}}},
  \bibinfo{author}{\bibfnamefont{R.}~\bibnamefont{{Keisler}}},
  \bibinfo{author}{\bibfnamefont{L.}~\bibnamefont{{Knox}}},
  \bibinfo{author}{\bibfnamefont{M.}~\bibnamefont{{Millea}}}, \bibnamefont{and}
  \bibinfo{author}{\bibfnamefont{C.}~\bibnamefont{{Reichardt}}},
  \bibinfo{journal}{\prd} \textbf{\bibinfo{volume}{87}}, \bibinfo{eid}{083008}
  (\bibinfo{year}{2013}), \eprint{1104.2333}.

\bibitem[{\citenamefont{Weinberg}(2008)}]{Weinberg:2008zzc}
\bibinfo{author}{\bibfnamefont{S.}~\bibnamefont{Weinberg}},
  \emph{\bibinfo{title}{{Cosmology}}} (\bibinfo{publisher}{OUP Oxford},
  \bibinfo{year}{2008}), ISBN \bibinfo{isbn}{9780191523601}.

\bibitem[{\citenamefont{{Rossi} et~al.}(2014)\citenamefont{{Rossi}, {Yeche},
  {Palanque-Delabrouille}, and {Lesgourgues}}}]{rossi/etal:2014}
\bibinfo{author}{\bibfnamefont{G.}~\bibnamefont{{Rossi}}},
  \bibinfo{author}{\bibfnamefont{C.}~\bibnamefont{{Yeche}}},
  \bibinfo{author}{\bibfnamefont{N.}~\bibnamefont{{Palanque-Delabrouille}}},
  \bibnamefont{and}
  \bibinfo{author}{\bibfnamefont{J.}~\bibnamefont{{Lesgourgues}}},
  \bibinfo{journal}{ArXiv e-prints}  (\bibinfo{year}{2014}),
  \eprint{1412.6763}.

\bibitem[{\citenamefont{Watanabe and Komatsu}(2006)}]{Komatsu2006}
\bibinfo{author}{\bibfnamefont{Y.}~\bibnamefont{Watanabe}} \bibnamefont{and}
  \bibinfo{author}{\bibfnamefont{E.}~\bibnamefont{Komatsu}},
  \bibinfo{journal}{Phys. Rev. D} \textbf{\bibinfo{volume}{73}},
  \bibinfo{pages}{123515} (\bibinfo{year}{2006}),
  \urlprefix\url{http://link.aps.org/doi/10.1103/PhysRevD.73.123515}.

\bibitem[{\citenamefont{{Hirata} and {Seljak}}(2005)}]{2005PhRvD..72h3501H}
\bibinfo{author}{\bibfnamefont{C.~M.} \bibnamefont{{Hirata}}} \bibnamefont{and}
  \bibinfo{author}{\bibfnamefont{U.}~\bibnamefont{{Seljak}}},
  \bibinfo{journal}{\prd} \textbf{\bibinfo{volume}{72}}, \bibinfo{eid}{083501}
  (\bibinfo{year}{2005}), \eprint{astro-ph/0503582}.

\bibitem[{\citenamefont{Maggiore}(2000)}]{Maggiore:1999vm}
\bibinfo{author}{\bibfnamefont{M.}~\bibnamefont{Maggiore}},
  \bibinfo{journal}{Phys.Rept.} \textbf{\bibinfo{volume}{331}},
  \bibinfo{pages}{283} (\bibinfo{year}{2000}), \eprint{gr-qc/9909001}.

\bibitem[{\citenamefont{Mangano et~al.}(2005)\citenamefont{Mangano, Miele,
  Pastor, Pinto, Pisanti et~al.}}]{Mangano:2005cc}
\bibinfo{author}{\bibfnamefont{G.}~\bibnamefont{Mangano}},
  \bibinfo{author}{\bibfnamefont{G.}~\bibnamefont{Miele}},
  \bibinfo{author}{\bibfnamefont{S.}~\bibnamefont{Pastor}},
  \bibinfo{author}{\bibfnamefont{T.}~\bibnamefont{Pinto}},
  \bibinfo{author}{\bibfnamefont{O.}~\bibnamefont{Pisanti}},
  \bibnamefont{et~al.}, \bibinfo{journal}{Nucl.Phys.}
  \textbf{\bibinfo{volume}{B729}}, \bibinfo{pages}{221} (\bibinfo{year}{2005}),
  \eprint{hep-ph/0506164}.

\bibitem[{\citenamefont{{Stewart} and
  {Brandenberger}}(2008)}]{Brandenberger2007}
\bibinfo{author}{\bibfnamefont{A.}~\bibnamefont{{Stewart}}} \bibnamefont{and}
  \bibinfo{author}{\bibfnamefont{R.}~\bibnamefont{{Brandenberger}}},
  \bibinfo{journal}{\jcap} \textbf{\bibinfo{volume}{8}}, \bibinfo{eid}{012}
  (\bibinfo{year}{2008}), \eprint{0711.4602}.

\bibitem[{\citenamefont{{Gerbino} et~al.}(2014)\citenamefont{{Gerbino},
  {Marchini}, {Pagano}, {Salvati}, {Di Valentino}, and
  {Melchiorri}}}]{Gerbino2014}
\bibinfo{author}{\bibfnamefont{M.}~\bibnamefont{{Gerbino}}},
  \bibinfo{author}{\bibfnamefont{A.}~\bibnamefont{{Marchini}}},
  \bibinfo{author}{\bibfnamefont{L.}~\bibnamefont{{Pagano}}},
  \bibinfo{author}{\bibfnamefont{L.}~\bibnamefont{{Salvati}}},
  \bibinfo{author}{\bibfnamefont{E.}~\bibnamefont{{Di Valentino}}},
  \bibnamefont{and}
  \bibinfo{author}{\bibfnamefont{A.}~\bibnamefont{{Melchiorri}}},
  \bibinfo{journal}{\prd} \textbf{\bibinfo{volume}{90}}, \bibinfo{eid}{047301}
  (\bibinfo{year}{2014}), \eprint{1403.5732}.

\bibitem[{\citenamefont{{{\"O}zsoy} et~al.}(2014)\citenamefont{{{\"O}zsoy},
  {Sinha}, and {Watson}}}]{ozoy/etal:2014}
\bibinfo{author}{\bibfnamefont{O.}~\bibnamefont{{{\"O}zsoy}}},
  \bibinfo{author}{\bibfnamefont{K.}~\bibnamefont{{Sinha}}}, \bibnamefont{and}
  \bibinfo{author}{\bibfnamefont{S.}~\bibnamefont{{Watson}}},
  \bibinfo{journal}{ArXiv e-prints}  (\bibinfo{year}{2014}),
  \eprint{1410.0016}.

\bibitem[{\citenamefont{{Lizarraga} et~al.}(2014)\citenamefont{{Lizarraga},
  {Urrestilla}, {Daverio}, {Hindmarsh}, {Kunz}, and
  {Liddle}}}]{2014arXiv1412.0407L}
\bibinfo{author}{\bibfnamefont{J.}~\bibnamefont{{Lizarraga}}},
  \bibinfo{author}{\bibfnamefont{J.}~\bibnamefont{{Urrestilla}}},
  \bibinfo{author}{\bibfnamefont{D.}~\bibnamefont{{Daverio}}},
  \bibinfo{author}{\bibfnamefont{M.}~\bibnamefont{{Hindmarsh}}},
  \bibinfo{author}{\bibfnamefont{M.}~\bibnamefont{{Kunz}}}, \bibnamefont{and}
  \bibinfo{author}{\bibfnamefont{A.~R.} \bibnamefont{{Liddle}}},
  \bibinfo{journal}{ArXiv e-prints}  (\bibinfo{year}{2014}),
  \eprint{1412.0407}.

\bibitem[{\citenamefont{Benson et~al.}(2014)}]{Benson:2014qhw}
\bibinfo{author}{\bibfnamefont{B.}~\bibnamefont{Benson}} \bibnamefont{et~al.}
  (\bibinfo{collaboration}{SPT-3G Collaboration}), \bibinfo{journal}{Proc.SPIE
  Int.Soc.Opt.Eng.} \textbf{\bibinfo{volume}{9153}}, \bibinfo{pages}{91531P}
  (\bibinfo{year}{2014}), \eprint{1407.2973}.

\bibitem[{\citenamefont{Isaacson}(1968)}]{Isaacson:1968zza}
\bibinfo{author}{\bibfnamefont{R.~A.} \bibnamefont{Isaacson}},
  \bibinfo{journal}{Phys.Rev.} \textbf{\bibinfo{volume}{166}},
  \bibinfo{pages}{1272} (\bibinfo{year}{1968}).

\bibitem[{\citenamefont{Maggiore}(2007)}]{maggiore2007}
\bibinfo{author}{\bibfnamefont{M.}~\bibnamefont{Maggiore}},
  \emph{\bibinfo{title}{Gravitational Waves}}, vol.~\bibinfo{volume}{1} of
  \emph{\bibinfo{series}{Gravitational Waves}} (\bibinfo{publisher}{OUP
  Oxford}, \bibinfo{year}{2007}), ISBN \bibinfo{isbn}{9780198570745}.

\end{thebibliography}

\end{document}